\documentclass[a4paper,12pt,preprint]{elsarticle}

\usepackage{hyperref}

\journal{SoftwareX}

\usepackage{hyperref}
\usepackage{graphicx,subcaption}
\usepackage{color}
\usepackage{placeins}
\usepackage{svg}
\usepackage{dcolumn}
\usepackage{bm}
\usepackage{mathtools} 
\usepackage{amssymb,amsfonts,amsmath}

\usepackage{ulem}

\usepackage[utf8]{inputenc}
\usepackage[T1]{fontenc}
\usepackage{float}

\newcommand{\AMBER}{AMBER}

\begin{document}

\begin{frontmatter}

\title{\AMBER: \textbf{A}lgorithm for \textbf{M}ultiplexing spectrometer \textbf{B}ackground \textbf{E}stimation with \textbf{R}otation-independence}


\author[psi]{Jakob Lass}
\author[sdsc]{Victor Cohen}
\author[psisdsc,sdsc]{Benjam\'in B\'ejar Haro}
\author[psi]{Daniel G. Mazzone}

\address[psi]{PSI Center for Neutron and Muon Sciences, 5232 Villigen PSI, Switzerland}
\address[sdsc]{Swiss Data Science Center, ETHZ, 8006 Zürich, Switzerland}
\address[psisdsc]{PSI Center for Scientific Computing, Theory and Data, Paul Scherrer Institut, 5232 Villigen PSI, Switzerland}
\cortext[coor]{\textit{Correspondence to}: Forschungsstrasse 111, 5232 Villigen PSI, Switzerland. \\ \textit{E-mail address}: jakob.lass@psi.ch}

\begin{abstract}
State-of-the art neutron spectrometers enable simultaneous measurements of high-dimensional datasets, allowing for a large collection rate of dynamic material properties. In this paper, we present the \textbf{A}lgorithm for \textbf{M}ultiplexing spectrometer \textbf{B}ackground \textbf{E}stimation with \textbf{R}otation-independence (\AMBER), which is a segmentation algorithm designed to decompose measured neutron scattering data into model-agnostic foreground and background contributions. The method takes advantage of the fact that background and foreground signals are measured simultaneously during the data collection process, relying on rotational independence of background contributions. The algorithm, initially developed for multiplexing neutron spectrometers
, aims to strongly reduce time consuming expert input, therefore promoting full data set usage while minimizing the source of systematic errors. 
\end{abstract}

\begin{keyword}
Inelastic Neutron Scattering \sep Machine Learning \sep Signal Decomposition \sep Background Determination 
\end{keyword}

\end{frontmatter}

\section*{Current code version}
\label{sec:CurrentCodeVersion}

\begin{table}[H]
\begin{tabular}{|l|p{6.3cm}|p{6.3cm}|}
\hline
\textbf{Nr.} & \textbf{Code metadata description} & \textbf{AMBER}  \\
\hline
C1 & Current code version & {\color{red}1.0.1}  \\
\hline
C2 & Permanent link to code/repository used for this code version & \url{https://gitea.psi.ch/lass_j/AMBER}  \\
\hline
C3 & Code Ocean compute capsule & N/A \\
\hline
C4 & Legal Code License   & Mozilla Public License 2.0 (MPL-2.0)  \\
\hline
C5 & Code versioning system used & git  \\
\hline
C6 & Software code languages, tools, and services used & Python, scipy, matplotlib, pytorch  \\
\hline
C7 & Compilation requirements, operating environments \& dependencies &  \\
\hline
C8 & If available Link to developer documentation/manual &  \url{https://AMBER-ds4ms.readthedocs.io/}  \\
\hline
C9 & Support email for questions & \url{jakob.lass@psi.ch}.com  \\
\hline
\end{tabular}
\caption{Code metadata (mandatory)}
\label{tab:Codemetadata} 
\end{table}


\section{Introduction}
\label{sec:intro}
Neutron scattering, alongside adequate modelling, enables a microscopic understanding of the prevailing interactions in a material. The experimental technique is utilized across a wide range of disciplines and material classes, and has been instrumental to many breakthroughs in fundamental solid state material questions \cite{Squires_2012,Boothroyd2020,TripleAxis}. The observed quantity measured in neutron experiments is the scattering probability $S(\vec{Q}, \omega)$, with $\vec{Q} = \vec{k}_i-\vec{k}_f$ and $\omega = E_i-E_f$ being the momentum and energy transfer between sample and neutron, respectively. The subscripts $i$ and $f$ denote the initial and final state of the neutron. $S(\vec{Q}, \omega)$ contains information of the microscopic Hamiltonian, encoding the materials' structure and its fundamental interactions. Understanding the Hamiltonian allows researchers to explain microscopic material properties and their behaviour under external stimuli such as magnetic field, pressure, or temperature, ultimately supporting industry to create novel devices for technical applications\cite{Boothroyd2020}. The removal of background features are not just needed when looking at coherent excitations but vital for methods calculating quantum mechanical entanglement\cite{Scheie2021}. In this paper we will focus on crystalline materials exhibiting coherent excitations, such as magnetic long-range ordered states with spin-wave spectra.

Traditionally, magnetic interactions have been probed with triple-axis neutron spectrometers \cite{TripleAxis}. Their working principle enables individual measurements of scattered neutrons at one specific energy and momentum transfer, using three independent angles to define $\vec{k}_i$, $\vec{k}_f$, $E_i$, and $E_f$. In contrast, multiplexing neutron instruments allow for simultaneous measurements of up to thousands of different points in $S(\vec{Q}, \omega)$. This is achieved by cross-normalizing multiple position-sensitive detectors, collecting neutrons over a quasi-continuous energy and wide angular range \cite{Groitl2016,Groitl2017,FlatCone,Lim2014,Lass2023CAMEA}. The instrument design reduces the data collection process to two independent angles defining the incident momentum ($\vec{k}_i$) and energy ($E_i$), while a spread of $\Delta k_f$ and $\Delta E_f$ is collected in a single acquisition. Alternative multiplexing neutron instruments are time of flight (ToF) spectrometers, in which the speed of neutrons is used to determine $\omega$ at many thousand scattering angles $\vec{k}_f$\cite{Boothroyd2020}. The data collection process in both instrument types involves a series of sample rotation scans, enabling the exploration of $S(\vec{Q}, \omega)$. 

\begin{figure}
    \centering
    \includegraphics[width=\linewidth]{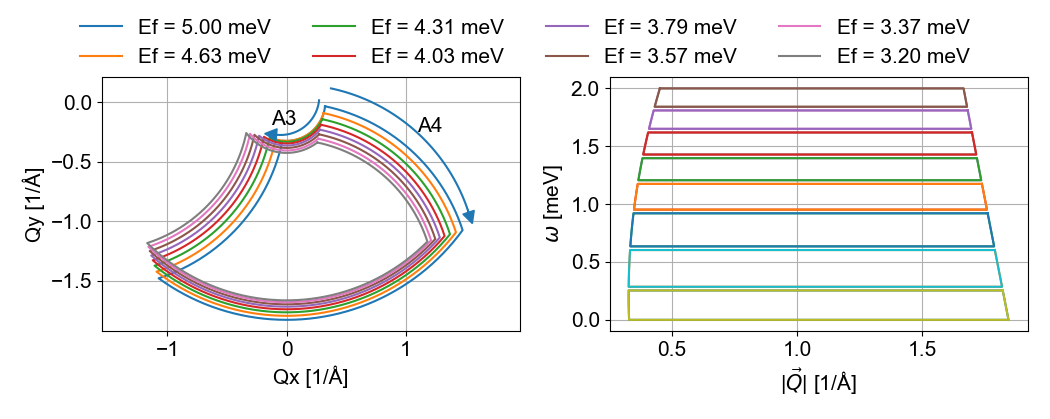}
    \caption{Coverage of a sample rotation scan on CAMEA. Left: Coverage in the two momentum transfer directions $Q_x$ and $Q_y$ for the eight final energies. The outlines are determined from the detector coverage, here -12$^\circ$ to 72$^\circ$ and the sample rotation, here from 0$^\circ$ to 90$^\circ$. Right: Cross-sectional cut through momentum $|\vec{Q}|$ = $\sqrt{Q_x^2+Q_y^2}$ and energy transfer $\omega$.}
    \label{fig:CAMEACoverage}
\end{figure}

In this manuscript we will highlight data stemming from the multiplexing CAMEA spectrometer at the Paul Scherrer Institut, Switzerland\cite{Lass2023CAMEA}, but we mention that \AMBER\ is applicable for multiple experimental disciplines and even datasets outside neutron scattering as long as they fulfil the requirements in sec.\ref{sec:limitations}. CAMEA is optimised for an efficient and detailed mapping of low-energy excitations under extreme experimental conditions, including sub-Kelvin temperature, high magnetic fields, and pressure. The instrument simultaneously collects neutrons over a quasi-continuous energy range of $\omega \sim$ 2 meV in an angular range of A$_4\sim$60 degrees (see Figure~\ref{fig:CAMEACoverage}). Larger energy and angular ranges are measured through consecutive instrument settings changing $E_i$. 
The simultaneous acquisition of sizeable $S(\vec{Q}, \omega)$ ranges generates challenges in the data curation process. Notably, it requires researchers to manually check for and mask out spurious background features. These spurious features are also present in triple-axis datasets, but they are more elusive due to the limited data coverage. Conventional analysis methods of neutron scattering data are based on narrow selections around candidate signals, discarding large parts in $S(\vec{Q}, \omega)$ space. The background is modelled for individual cuts using low-order polynomials which are sometimes extrapolated far from the original $S(\vec{Q}, \omega)$ region. The classical process is time consuming, requires expert input, and does not exploit the full information contained within the dataset. It is conceivable that important features and novel phenomena have previously been overlooked as relevant sectors have remained unanalysed. The automatic signal decomposition algorithm presented here gives access to objective data curation strategies at a much quicker rate, and allow for scattering models with an objective separation of background and foreground contributions. Ultimately, this enables a more accurate comparison between experimental results and theoretical predictions. 

An alternative to both AMBER and the above described masking method is to prepare the experimental system in a setup where background is directly measured. However, this cannot always be achieved, either due to the sample properties, the lack of availability beam time or the sample environment utilized. However, such backgrounds are expected to be similarly good or better than a background estimate from AMBER. An alternative method often used in ToF experiments is a measurement of an empty sample holder. However, both approaches are time consuming.

\section{Software description}
\subsection{Physical Assumptions}\label{sec:PhysicalAssumptions}
Neutron instruments remain in fixed positions during acquisition periods to collect scattered neutrons at desired points in ($\vec{Q}$, $\omega$). Standard triple-axis spectrometers consecutively acquire a series of points by changing $\vec{Q}$ or $\omega$,  adjusting most of the angles defining the energy and direction of the incoming and outgoing neutrons. Multiplexing spectrometers cover a range of energy transfers and scattering vectors in a single acquisition, collecting a large portion of $S(\vec{Q}, \omega)$ in a series of sample rotations angles, denoted A$_3$ scans. We assume the sample to be a single crystal, possessing $S(\vec{Q}, \omega)$ signals that depend on the sample rotation. The background contributions are assumed to be A$_3$ position independent, but depend on the length of the scattering vector $|\vec{Q}|$ and the energy transfer $\omega$. This assumption is justified as parasitic signals mainly originate from neutrons scattered by the large sample environments containing the sample or from air scattering along the neutron path. While this signal is difficult to predict but remains constant for fixed instrument positions. Therefore, the accidental scattering depends only on ($|\vec{Q}|,\omega$). Further, most background contributions are empirically known to be continuous and smooth and can be described by a smoothly varying function in $|\vec{Q}|$ and $\omega$. 

\subsection{Mathematical Formulation}
We consider neutron scattering data collected in the scattering plane orthogonal to the A$_3$ rotation axes, amounting to a three-dimensional $S(\vec{Q}, \omega)$ dataset. We assume that the $\vec{Q}$-plane and energy axis are discretized into a constant sized grid with $n_x$, $n_y$ and $n_{\omega}$ bins along the orthogonal $Q_x$, $Q_y$ and $\omega$ directions, respectively. We introduce a discretization of the polar $Q_x$, $Q_y$ coordinates and define by $r_{\mathrm{max}}:= |\vec{Q}_{\mathrm{max}}|$ the maximal radius of the measurement data. The radial line $(0,r_{\mathrm{max}})$ is discretized into $n_q$ grid points. The signal decomposition problem can then be stated as follows: Given a set of observed points $Y$, i.e. scattering intensities, at specified $\vec{Q}, \omega$ points in $\mathbb{R}^{n_x\times n_y\times n_{\omega}}$, find the signal $X$ in $\mathbb{R}^{n_x\times n_y\times n_{\omega}}$ and the background $B$ in $\in\mathbb{R}^{n_x\times n_y\times n_{\omega}}$ such that $Y = X + B$. 
To drive the signal separation, one needs to leverage prior knowledge on the background and the signal.
The background $B$ is assumed invariant under rotation around the origin corresponding to A$_3$ in the $\vec{Q}$-plane and smoothly changing along $|\vec{Q}|$. Thus, it is rewritten as $B = \mathcal{R}b$  where $\mathcal{R}$ is an operator rotating a vector $b$ belonging to $\mathbb{R}^{n_q\times n_{\omega}}$, i.e., $\mathcal{R}\colon b \in \mathbb{R}^{n_q \times n_{\omega}} \mapsto \mathcal{R}b \in \mathbb{R}^{n_x \times n_y \times n_{\omega}}$ such that $(\mathcal{R}b)_{x,y,e} = b_{r,e}$ for all triplets $(x,y,r)$ with $x^2 + y^2 = r^2$.
Moreover, one can assume that the reconstructed signal $X$ is sparse in the entire volume and smooth along the energy axis. 

Hence, we formulate the signal separation as the following the minimization problem:
\begin{align}
    \min_{X,b} &\frac{1}{2}\lVert Y-X-\mathcal{R}b\rVert_{2}^2+\lambda\vert| X |\vert_{1} +\frac{\beta}{2} \mathrm{Tr} \left( b^T L_{b} b \right) +\frac{\mu}{2} \boldsymbol{1}_{n_x}^TX^T L_{\omega} X\boldsymbol{1}_{n_y} \label{pb:minimization}
\end{align}
The first term of Eq.~\eqref{pb:minimization} is the fitting term ensuring the signal-noise decomposition. The second term imposes sparsity of the signal features.
The third and fourth terms are Laplacian regularizations~\citep{Tong2006} enforcing smoothness on the background and signal. 
$\lambda$, $\beta$ and $\mu$ are real positive scaling factors. $L_b$ and $L_{\omega}$ are Graph Laplacians belonging respectively to $\mathbb{R}^{n_q \times n_q}$ and $\mathbb{R}^{n_{\omega} \times n_{\omega}}$. We note that the third and fourth terms can be equivalently written as 
\begin{align}
    \mathrm{Tr}(b^T L_b b) &= \sum_{j=1}^{n_{\omega}} \sum_{i=1}^{n_{q}} \left(b_{i+1,j} - b_{i,j}\right)^2\\
    \boldsymbol{1}_{n_x}^TX^T L_{\omega} X \boldsymbol{1}_{n_y} &= \sum_{i,j=1}^{n_{x},n_y}\sum_{k=1}^{n_{\omega}} \left(X_{i,j,k+1} - X_{i,j,k}\right)^2,
\end{align}
where $X_{i,j,k}$ is the ijk-th element of X and $b_{r,k}$ is the rk-th element of $b$. 
The graph Laplacian matrices $L_b$ and $L_{\omega}$ share a common structure corresponding to the Laplacian of a 1D chain. They are given by a tri-diagonal matrix with the vectors $\left(-1,1,1,\cdots,1,1,-1\right)$, $\left(1,2,2,\cdots,2,2,1\right)$, and $\left(-1,1,1,\cdots,1,1,-1\right)$. For instance, the Laplacian matrix of a chain containing $5$ vertices is;
\begin{align*}
    L = 
    \begin{pmatrix}
        1 & -1 & 0 & 0 & 0 \\
        - 1 & 2 & 1 & 0 & 0\\
        0 & 1 & 2 & 1 & 0 \\
        0 & 0& 1  & 2 & -1 \\
        0 & 0 & 0  & -1 & 1 
    \end{pmatrix}.
\end{align*}

\subsection{Coordinate descent algorithm to solve the minimization problem}
\label{sec:solution}
The loss function of Eq. \ref{pb:minimization} for a given signal $X$ and background $b$ is denoted by $\mathcal{L}(X,b)$ . 
Equation.~\ref{pb:minimization} can in principle be solved using proximal gradient based algorithm but a significant number of iterations may be required to reach an optimal solution. 
Instead, we use a coordinate descent algorithm, in which $\mathcal{L}$ is iteratively minimized with respect to $X$ keeping $b$ fixed and with respect to $b$ keeping $X$ fixed. This method enables to use closed-form updates.
Given initial values of the signal $X^0$ and the background and $b^0$, we iterate $K$ times for $k=0,\ldots,K-1$,
\begin{align}
    & X^{k+1} \leftarrow \mathrm{argmin}_{X} \mathcal{L} \left(X, b^k \right) = \left(\boldsymbol{I} +\mu  L_{\omega} \right)^{-1} S_{\lambda}\left( Y - \mathcal{R} b\right) \label{eq:coordinate_descent_Xupdate} \\
    & b^{k+1} \leftarrow \mathrm{argmin}_{b} \mathcal{L} \left(X^{k+1}, b \right)=  \big( \Gamma + \beta L_{q} \big)^{-1} \mathcal{R}^*( Y -  X). \label{eq:coordinate_descent_Bupdate}
\end{align} 
where $\Gamma$ is a $n_q \times n_q$ diagonal matrix whose elements are the number of measurements in each radial bin, $\mathcal{R}^*$ is the adjoint of operator $\mathcal{R}$ (i.e., $\langle \mathcal{R}U , V  \rangle = \langle U, \mathcal{R}^* V\rangle$ where $\langle, \rangle$ is the standard inner product in Euclidean space) and $S_{u}(x) = \mathrm{sign}(x) \max(\vert x \vert -u, 0)$. 
One can show that $\mathcal{R}^*$ maps any $n_x \times n_y \times n_{\omega}$ matrix into a $n_q \times n_{\omega}$ matrix, whose elements are the circular sum across angles for each radial bin (see \ref{App:Proof} for the detailed proof.) 
The main advantage of such an approach is that both subproblems (Eqs. ~\eqref{eq:coordinate_descent_Xupdate} and ~\eqref{eq:coordinate_descent_Bupdate}) have closed-form solutions.
One can observe that the inverse of the matrices $\boldsymbol{I} +\mu  L_{\omega}$ and $\Gamma +\beta  L_{q}$ do not depend on the data and they can be precomputed ahead of time.

\subsection{Hyperparameter tuning}
\label{sub:hyperparam}

Equation~\ref{pb:minimization} depends on a set of parameters $\theta = \left( \lambda, \beta, \mu \right)$, which needs to be tuned to find an optimal solution. 
Each plays a specific role in the signal-background decomposition. $\lambda$ and $\mu$ impact the signal sparsity and the signal smoothness along the energy axis, $\beta$ controls the background smoothness along the radial axis (in polar coordinates of the Q-plane). 

\subsubsection{Heuristic Parameter Determination}\label{sec:parameters}
The hyperparameter space exceeds the limit for a naive optimization without estimators of their values. Thus, an initial, heuristic method is used to choose these values. We propose the following
\begin{itemize}
    \item $\lambda$: Given the sparsity assumption, it is natural to consider the Median Absolute Deviation (MAD) which is a robust estimator of the standard deviation. We set $\lambda\colon =  k \mathrm{MAD}(Y)$ where $k=1.468$ is a scaling factor allowing the robust estimation of the standard deviation of the data~\citep{Croux1993}. 
    \item $\mu$: This parameter penalises the intensity variations along the energy axis. Its order of magnitude can be estimated by the magnitude of the variance of $Y$ along the energy axis, i.e., $\mu:= \mathrm{Mean}\left(\mathrm{Var}_{e}(Y) \right)$.  
    \item $\beta$: This parameter corresponds to a smoothing penalty on the background. It can be estimated through a cross-validation procedure including two steps. Data points containing mainly signal are removed through a thresholding, leaving points assumed to contain mostly background, i.e. $X \approx 0$. This results in $B\sim Y$ and thus one uses the RMS of $Y-B$ as a measure. The thresholding consists of masking values that are above a given quantile, $q$, of the observed values $Y$. 
\end{itemize}
The approach provides a generic initial setting of the hyperparameters. However, the strategy can be suboptimal depending on the acquired measurements and the sample. It is also mentioned that the determination of $\beta$ is potentially time consuming depending on the range of possible values. 
Hence, hyperparameter tuning can be improved based on instrument and spectrum knowledge (smoothness, sparsity, etc.). However, the computation time of our denoising algorithm is low - on the order of 1 minute - allowing the user to evaluate the signal decomposition for a wide range of hyperparameters.
To check the applicability of AMBER on a given dataset it is suggested to perform the following steps. Firstly, identify the signal to background level, i.e. estimate the percentage of data which contain only background. Next, estimate both $\lambda$ and $\mu$ as described above and find $\beta$ through a cross-validation. This provides a starting point to perform the first background subtraction. In cases where AMBER does not perform satisfactory, as exemplified in \ref{sec:limitations}, the subtracted data will have substantial regions of negative intensity or artefacts from abrupt changes in the background estimation. These effects signify cases in which the heuristic to determine the hyperparameters leads to less accurate background estimates. However, it is not excluded that a set of parameters would result in a satisfactory background estimation.

\subsection{Code implementation}
\AMBER\ has been implemented in a python library, called \textit{AMBER-ds4ms} relying only on well-established and widely-used packages, namely numpy~\cite{numpy} Scipy~\cite{Scipy}, and Matplotlib~\cite{Matplotlib}. It is compatible with all recent versions of python and tested against versions 3.9 through 3.13. The documentation is hosted at ReadTheDocs~\cite{documentation}. 

As an example, the latest MJOLNIR\cite{Lass2020MJOLNIR} version (1.3.5) contains a dedicate background object through which the \AMBER\ decomposition algorithm can be performed. MJOLNIR has been written  for a wide range multiplexing spectrometers and can thus be used in many neutron scattering experiments. In particular, data is binned along $Q_x$, $Q_y$ and $\omega$. The decomposition process is semi-automatised with standard values for the hyperparameters, but some tuning may be required. Firstly, the binning parameters $n_x, n_y, n_{\omega}$ and $n_q$ are chosen. Secondly, the algorithm hyperparameters $(\lambda,\beta,\mu)$ can be either set manually based on expert knowledge or using the heuristic approach mentioned in Sec.~\ref{sec:parameters}.

\subsection{Limitations}\label{sec:limitations}
The algorithm relies on a set of assumptions, performing suboptimal in cases in which they are not fulfilled. Here we explicitly list these assumptions and highlight common neutron scattering situations in which these are violated:
\begin{enumerate}
    \item Rotation independence of the background.
    \item Smooth change of background along energy and $|\vec{Q}|$.
    \item Sparse but continuous signal in energy and $|\vec{Q}|$.
\end{enumerate}
The elastic line - the entire scattering plane at 0 energy and energy transfer within the instrument resolution - is a natural region in which all assumptions are violated. More specifically, assumption 1 is invalid as sample holders usually have texture or a non-isotropically shape. The abrupt change of intensity between elastic and inelastic scattering invalidates assumption 2 while Bragg peaks invalidates the smoothness and continuity of the signal in energy and $|\vec{Q}|$. Accidental Bragg scattering, denoted as Currat-Axe Spurions, \cite{TripleAxis,Lass2023CAMEA} also violate the assumptions, in particular assumption 3. Lastly, for samples where $S(\vec{Q}, \omega)$ is dominated by a constant, scattering vector-independent signal, e.g. potentially the top of a magnon band or crystal electric fields, challenges assumption 3. We mention that in cases of strongly absorbing samples or a highly irregular sample holders, the rotation independence is no longer valid, even deeply inelastically. An absorption correction can be applied prior to the background model while the sample holder case cannot be treated with AMBER.

We further note that AMBER is also applicable for low-dimensional signals, in which the neutron scattering signal along some crystallographic directions remain unchanged. The main requirement for \AMBER\ to perform well is the presence of sufficient background volume at all constant energy planes.

\section{Numerical experiment} 
Neutron scattering signals stemming from magnetic materials contain a multitude of different characteristics and shapes. In long-range ordered magnets, the most prominent signals are vibrations of the magnetic moments (spin waves). They are usually well-defined circular objects located in momentum space with a conical expansion along the energy \cite{Squires_2012,Boothroyd2020,Yamani2010}. In other cases, such as in frustrated systems and materials exhibiting strong quantum fluctuations, the signals are often broadened in energy and $\vec{Q}$ \cite{Enderle2010, Sibille2020, Halloran2025}. For demonstration purposes, we limit our examples to spin waves, and mention that the algorithm should be tested on a case to case basis for excitations belonging to the latter case. Two benchmark cases utilizing a well-characterized MnF$_2$ single crystal sample are presented in appendices \ref{App:SyntheticMnF2} and \ref{App:MnF2}; one using synthetic data and the other high quality neutron scattering data. In the case of synthetic data, \AMBER\ succeeds in separating signal and background and outperforms the basic median approach across all levels of signal to noise. For the real world data, \AMBER\ performs on par with the method of background subtraction through masking, where the optimised mask is known from the dispersion relation of MnF$_2$. In the main manuscript, we present the case of a mosaic of VOSe$_2$O$_5$ single crystals. This example serves as a litmus-test of our method in terms of data quality and trade-off between computational time to manual effort.

\begin{figure}
    \centering
    \begin{subfigure}[t]{.49\linewidth}
      \centering
      \includegraphics[width=\linewidth]{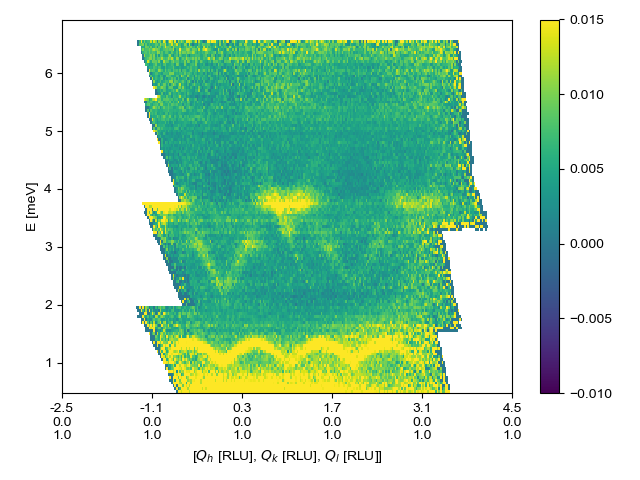}
      \caption{}
      \label{fig:2DData}
    \end{subfigure}\\
    \begin{subfigure}[t]{.49\linewidth}
      \centering
      \includegraphics[width=\linewidth]{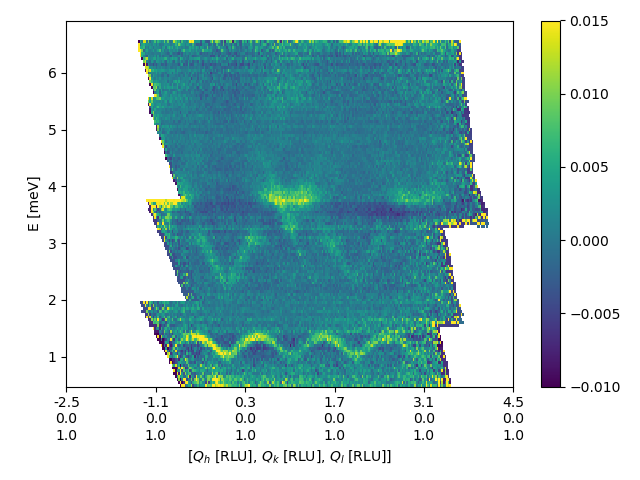}
      \caption{}
      \label{fig:2DAMBER}
    \end{subfigure}
    \begin{subfigure}[t]{.49\linewidth}
      \centering
      \includegraphics[width=\linewidth]{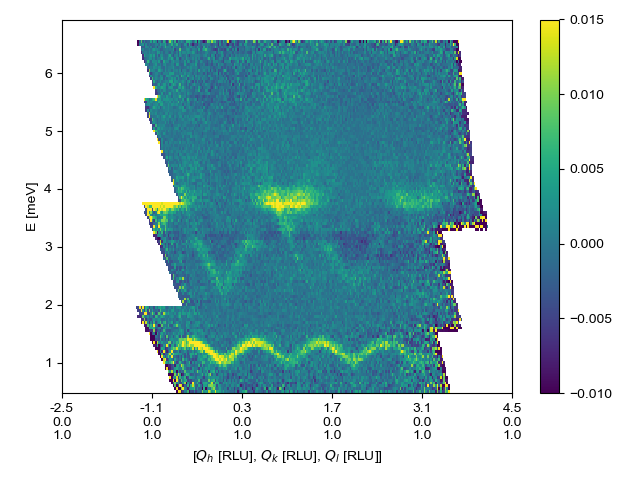}
      \caption{}
      \label{fig:2DExpert}
    \end{subfigure}
    \caption{Scattering intensities of VOSe$_2$O$_5$ as a function of energy transfer and scattering vector along ($H$,0,1) showing (\textbf{a}) the raw data, (\textbf{b}) subtracted data using \AMBER\ and (\textbf{c}) the results from an expert performing a manual subtraction. All plots were performed using an integration of 0.56 Å orthogonal to the cut, 0.014 Å along the cut, and 0.05 meV along the energy axis. }\label{fig:2DDataOverview}
\end{figure}

The data were acquired using CAMEA with instrument settings tabulated in Tab.~\ref{tab:VOSe2O5CAMEA}. The sample was measured within the horizontal ($H$,0,$L$) plane. The data reveal a low signal-to-noise level and a complex excitation spectrum shown in Fig. Fig.~\ref{fig:2DDataOverview}(a). It contains a series of spin waves extending up to 6.5 meV. While the magnetic excitations disperse mainly along the crystallographic $H$ direction, the $L$ dependence is limited. Thus, the data is presented in cuts along $H$ and energy transfer. The four strongest sample signatures are the three spin waves dispersing between 1 and 1.6 meV and 2.2 and 3 meV, and a less dispersive one around 3.7 meV. The lowest mode is visible across the entire $H$ range, revealing minima at integer values of $H$. The second magnon features equal minima, but has a modulated intensity. The higher mode is only visible around odd values of $H$, and is superimposed with a flat mode at 3.7 meV. Only data above the elastic line, i.e. $\omega$ $>$ 0.5 meV and without spurions are shown.

\begin{table}[!ht]
    \centering
    \begin{tabular}{c|c|c|c|c}
       Energy [meV]  &  A$_3$ Range [deg] & A$_3$ Steps & A$_4$ [deg] & Time/Step [s]  \\ \hline
       5.0  & -28 - 62 & 91 & -48, -52, -68, -72 & 60.5 \\
       5.13 & -28 - 62 & 91 & -48, -52, -68, -72 & 61.8 \\
       6.8 & -28 - 62 & 91 & -41, -45, -61, -75 & 65.2 \\
       6.93 & -35 - 55 & 91 & -41, -45, -61, -65 & 64.6 \\
       8.54 & -42 - 48 & 91 & -36, -40, -56, -60 & 70.5 \\
       8.60 & -42 - 48 & 91 & -36, -40, -56, -60 & 69.9 \\
       8.67 & -42 - 48 & 91 & -36, -40, -56, -60 & 70.2 \\
       8.73 & -42 - 48 & 91 & -36, -40, -56, -60 & 70.8 \\
       9.54 & -84 - 46 & 131 & -36, -40 & 83.1 \\
       9.60 & -84 - 46 & 131 & -36, -40 & 83.3 \\
       9.67 & -84 - 46 & 131 & -36, -40 & 85.0 \\
       9.73 & -84 - 46 & 131 & -36, -40 & 85.5 \\
    \end{tabular}
    \caption{Instrument settings for the experimental VOSe$_2$O$_5$ data. For each combination of energy and A$_4$, for e.g. an incoming energy of 6.8 meV a total of four A$_3$ scans have been performed. Thus, the data set contains 40 scans in total.}
    \label{tab:VOSe2O5CAMEA}
\end{table}

We compare the performance of \AMBER\ against an expert background estimation, stemming from a manual mask of the foreground signals. This roughly 8 hours work allows to find the background signal of each data file by averaging along constant $|\vec{Q}|$ and is subtracted from the data. This is a common method already used in other studies\cite{Facheris2022,Sala2023}. The results for VOSe$_2$O$_5$ are shown in  Fig.~\ref{fig:2DDataOverview}(c).  \AMBER\ was run with a binning $Q_x=Q_y=$0.03 Å$^{-1}$ and $E =$0.05 meV. $\lambda$ and $\mu$ were found according to sec.~\ref{sec:parameters}, and $\beta$ through the cross validation method with a background threshold of 60 \%. The final RMSE for this cross validation is 107.98, which required a validation of \textbf{\textit{28}} iterations resulting in a computational time of 17 min (each iteration took $\sim$ 50 s). The results presented in Fig.~\ref{fig:2DDataOverview}(b) were obtained with the parameters shown in Tab.~\ref{tab:VOSe2O5}. 
\begin{table}[!ht]
    \centering
    \begin{tabular}{l|c|c|c}
       Parameter  & $\lambda$ & $\beta$ & $\mu$ \\\hline
       Value  & 0.00775 & 347.97 & 0.00170
    \end{tabular}
    \caption{Parameters used for the \AMBER\ background of VOSe$_2$O$_5$ shown in Fig.~\ref{fig:2DAMBER}.}
    \label{tab:VOSe2O5}
\end{table}

\begin{figure}
    \centering
    \begin{subfigure}[t]{.49\linewidth}
      \centering
      \includegraphics[width=\linewidth]{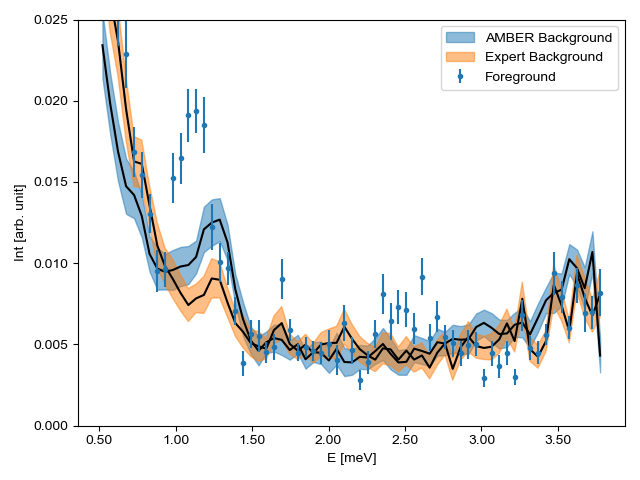}
      \caption{}
      \label{fig:1D_001}
    \end{subfigure}
    \begin{subfigure}[t]{.49\linewidth}
      \centering
      \includegraphics[width=\linewidth]{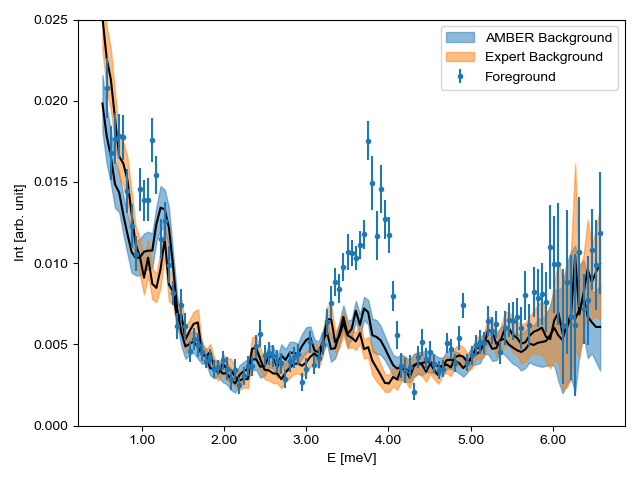}
      \caption{}
      \label{fig:1D_101}
    \end{subfigure}\\
    \begin{subfigure}[t]{.49\linewidth}
      \centering
      \includegraphics[width=\linewidth]{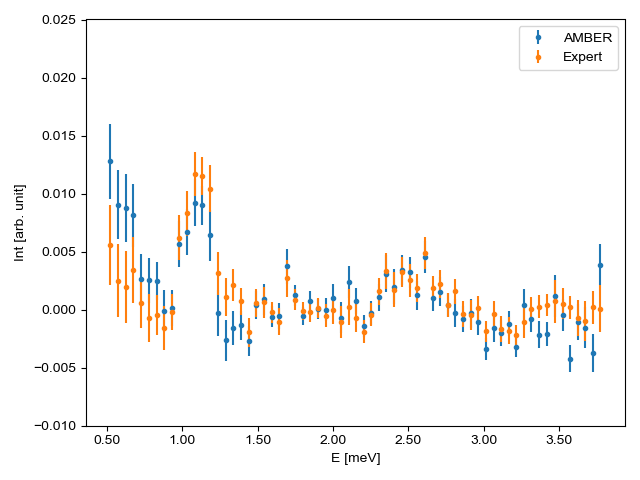}
      \caption{}
      \label{fig:1D_001_sub}
    \end{subfigure}
    \begin{subfigure}[t]{.49\linewidth}
      \centering
      \includegraphics[width=\linewidth]{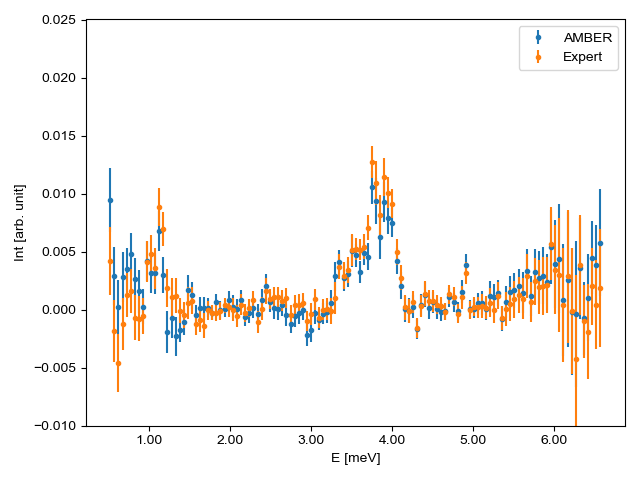}
      \caption{}
      \label{fig:1D_101_sub}
    \end{subfigure}\\%
    \caption{Scattering intensity of VOSe$_2$O$_5$ as function of energy transfer at (0,0,1) ((\textbf{a}) and (\textbf{c})) and at (1,0,1) ((\textbf{b}) and (\textbf{d})). These are shown with the unsubtracted data overplotted with the expert and \AMBER\ backgrounds shown as a solid line with shaded 1$\sigma$ error estimates ((\textbf{a}) and (\textbf{b})) and the subtracted data using the two methods ((\textbf{c}) and (\textbf{d})). All cuts were performed with a $Q$ integration of 0.1 Å$^{-1}$ and 0.05 meV steps along the energy axis.}
    \label{fig:1DData}
\end{figure}


From Fig.~\ref{fig:2DDataOverview}(b) and (c) it is seen that \AMBER\ produces acceptable results close to ground truth in less than one minute, without expert knowledge. As 2D intensities are difficult to compare, 2 1D cuts along energy have been prepared in Fig.~\ref{fig:1DData}. (a) and (b) show the data with the background estimates while (c) and (d) show the subtracted data. It can, however, be seen that some regions suffer from over-subtraction, i.e. location where the background intensity is overestimated. They appear around the low energy magnon branch (1 - 1.5 meV) and for the the mode around 3.5 - 4 meV. It is here emphasized that such over-subtraction also occurs in the expert background but to a lesser degree. 

Improvements to the background procedure can be achieved by extending the parameters from scalar values energy transfer dependent variables (see App.~\ref{App:ParameterExtension}).

\FloatBarrier{}

\section{Impact}

The data collection strategy of modern neutron spectrometers such as the multiplexing instrument CAMEA allow for a simultaneous measurement of foreground and background features. The current best-practice methodology for background determination  in already measured data is to manually mask areas containing foreground features, and use the remaining data as background. An interpolation among background points enables an estimation of the background below the signal taking advantage of the sample rotation independence (see Sec.~\ref{sec:PhysicalAssumptions}). However, this process is labour intensive and requires expert-knowledge about the expected signal  and about the instrument. The process is also highly volatile to subjective errors.  \AMBER\ allows non-expert users to perform an efficient foreground-background segmentation of three-dimensional data sets obtained on multiplexing spectrometer. Using \AMBER\, the background determination is reduced  to minutes, it is objective and reproducible. This is especially practical for measurements suffering from  low signal-to-noise ratios. However, we advice users not to blindly trust the algorithm output.

A logical expansion path of \AMBER\ is to include time-of-flight neutron spectroscopy data, which is also based on rotation scans following the same assumptions as described in sec.~\ref{sec:PhysicalAssumptions}. Since the standard time-of-flight analysis software, Mantid\cite{Mantid14}, is build on Python, we are positive that an implementation of \AMBER\ is achievable with relatively low efforts. \AMBER\ may also be beneficial for x-ray and electron based experimental techniques, such as angle-resolved photoemission, resonant inelastic x-ray scattering, tunnelling electron spectroscopy and other techniques using large area detectors. The mathematical formulation behind the algorithm extends to a third dimension of $\vec{Q}$ either through an expansion of the found background estimate from 2 to 3D, i.e. being dependent of $|\vec{Q}|_{x,y}$ calculated from the in-plane scattering vector, the out of plane scattering vector and energy. Alternatively, one could include $\vec{Q}_z$ directly in the calculation of $|\vec{Q}|$ or ignore it altogether. Both of these suggestions impose assumptions not based on known physical background behaviour. A far greater issue is that of computational complexity. The computational complexity of one iteration step would significantly increase due to the matrix inversion. Alternatively, if the algorithm is made tractable it could be distributed across a multi-kernel setup like a cluster or a graphical unit. 

\section{Conclusion}
In this article we have shown how background and signal contributions from multiplexing neutron scattering instruments can be segmented using \AMBER\ under the assumption that the background arises from rotational independent background features. Our method is comparable to the expert performance, but superior in time efficiency. The effectiveness of the algorithm has been tested on synthetic and real data, show in app.\ref{App:SyntheticMnF2} and \ref{App:MnF2}, as well as by comparing to an expert background in the case of VOSe$_2$O$_5$. \AMBER\ faster than manually performed background decomposition, and allows non-expert users to generate an objective background estimation.

A next step for the algorithm is to include Poisson statistics, because of the nature of the counting statistics. This would be a more natural approach for uncertainty estimation. In turn,  this could lead to an estimation of background uncertainty highlighting regions where more statistic is required. One could leverage this to guide experiments by optimization of time allocation.

\section*{Conflict of Interest}
We confirm that there are no conflicts of interest associated with this publication.

\section*{Funding}

This research was funded by the Swiss Data Science Center through the "Data Science for Multiplexing Spectrometers (DS4MS)" grant no. C12-16L, funded in the fifth call under the LSI track.

\section*{Acknowledgements}
We thank Dr. Jonathan White for contributing the VOSe$_2$O$_5$ data for this study. 


\bibliography{bibliography}
\newpage
\appendix

\section{Benchmarking}\label{App:Benchmarking}

We used the well-know material MnF$_2$\cite{Yamani2010,Morano2024} to benchmark \AMBER\ in two different setups. 
Firstly, we used a synthetic data set based on the analytical theoretical model of MnF$_2$ for a parametric study of our method's robustness towards low signal-to-noise levels. Secondly, a real world data set of a MnF$_2$ single crystal measured at CAMEA was used to confirm the applicability of our method. 

\subsection{Synthetic data - \texorpdfstring{MnF$_2$}{MnF2}}\label{App:SyntheticMnF2} 
The initial test for \AMBER\ was to segment background and foreground contributions in an artificial dataset. Since the method is intended to be used on localized and sparse inelastic signals, the arc-typical setup is a well-defined magnon spectrum, e.g. that of MnF$_2$. Synthetic data further allows full control over the relative intensities of signal (S) and background (N), i.e. S/N ratio. 
    
In a general case, magnon spectra are calculated through numerical methods using linear spin-wave theory\cite{Toth2015,Boothroyd2020}, but a few systems can be (approximately) described by analytical functions. This is the case for MnF$_2$ exhibiting a single magnon branch parametrized through \cite{Yamani2010}
\begin{align}
  \omega^2(H,K,L) = \left(2Sz_2J_2+D+2Sz_1J_1\sin[2]{\pi Q_L)}\right)^2- \label{eq:theoryMnF2}\\
  \left(2Sz_2J_2\cos{\pi Q_H}\cos{\pi Q_K}\cos[2]{\pi Q_L}\right)^2 , \notag
\end{align}
where $S$ is the manganese spin of 5/2, and $z_1 = 2$ and $z_2 = 8$ are the numbers of nearest and next-nearest neighbours, respectively. In many cases, the parameters of interest are the directions and magnitude of the couplings, which in this example are two magnetic coupling strengths $J_1$ and $J_2$ as well as the magnetic anisotropy along the crystallographic z direction, $D$. Equation~\eqref{eq:theoryMnF2} only predicts the spin-wave dispersion, i.e. its relation between the scattering vector $\Vec{Q}$ and its energy, but not its intensity. For the present use-case we set the intensity equal across the dispersion. The calculation further describes an infinitely thin dispersion, neglecting the instrument response function. The instrument broadening occurs due to finite size instrument components alongside trade-offs between narrow instrument resolution and neutron flux\cite{TripleAxis,Boothroyd2020}. In the present case, the resolution function was crudely approximated by a constant Gaussian with a width of $\sigma = 0.25$ meV. The instrument background was imitated by an intensity function depending only on the scattering angle created from one Lorentzian and a Gaussian function. Thus, it only depends on $|Q|$. The first component mimics the direct beam contribution, i.e. small angle scattering originating around the sample and the sample environment, while the second peak reflects the general increase in background for large scattering angles. A flat background was added to simulate the electronic noise from the instrumental setup. This contribution is kept constant through the benchmarking process. Their combined expression is given by
\begin{equation}
    20A\left(\frac{\gamma}{A_4^2+\gamma^2}+0.02\ \mathrm{e}^{-\frac{\left(A_4+120\right)^2}{2\ \sigma^2}}\right)+0.01,
\end{equation}
where $\gamma$ = 10 and $\sigma$ = 20. These values and the relative intensities of the two functions were found to mimic best experimental data. The counting statistic was included by drawing from the signal and background combination using a Poisson distribution. These raw counts were treated identically to experimental data, i.e. rescaled through the instrument normalization and measurement time (see Ref. \cite{Lass2020MJOLNIR} for details).
    



\begin{figure}
    \centering
    \begin{subfigure}{.5\textwidth}
      \centering
      \includegraphics[width=\linewidth]{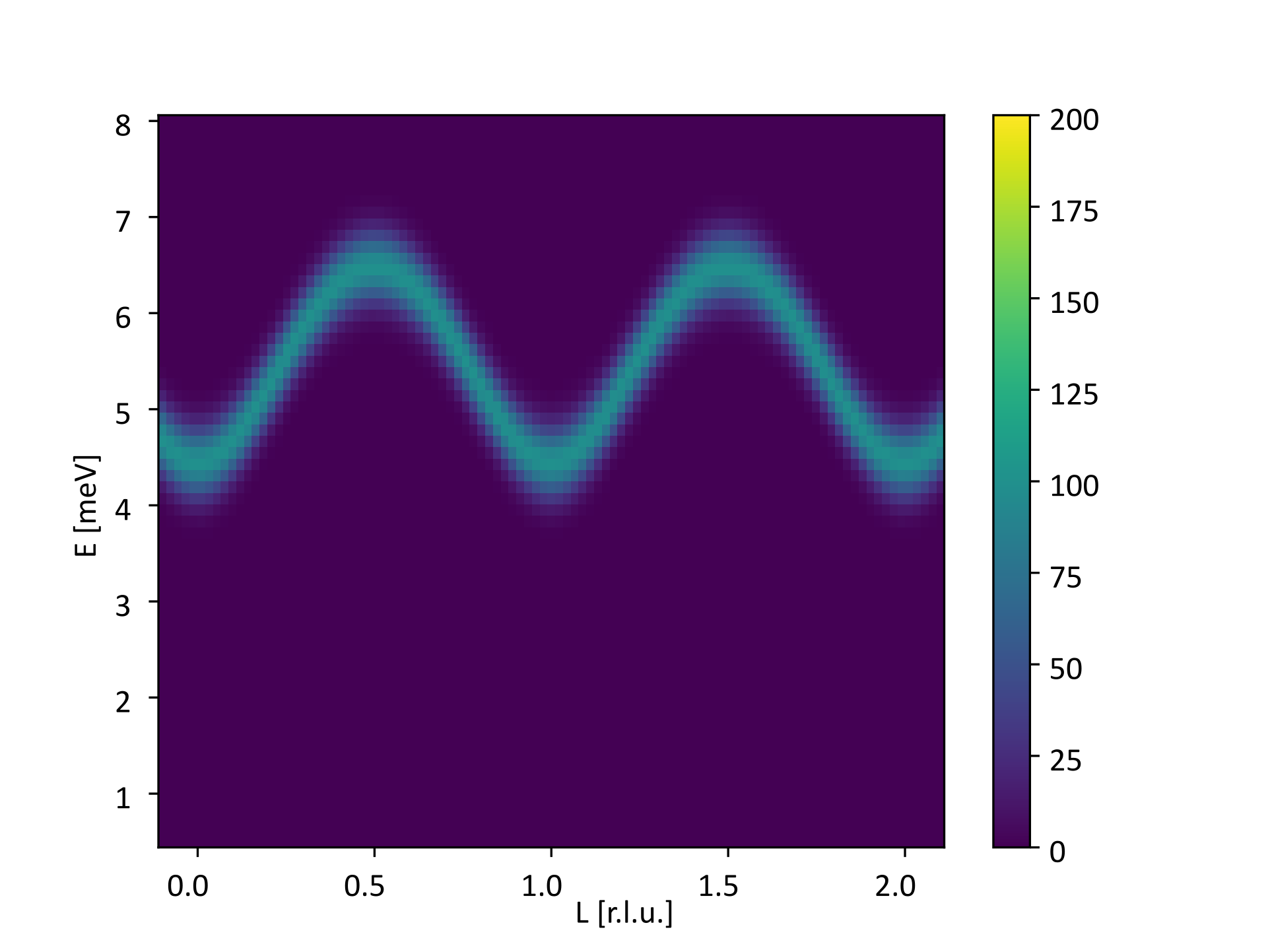}
      \caption{QvE plot ($H=0.25$)}
      \label{fig:gt_data:QvE}
    \end{subfigure}%
    \begin{subfigure}{.5\textwidth}
      \centering
      \includegraphics[width=\linewidth]{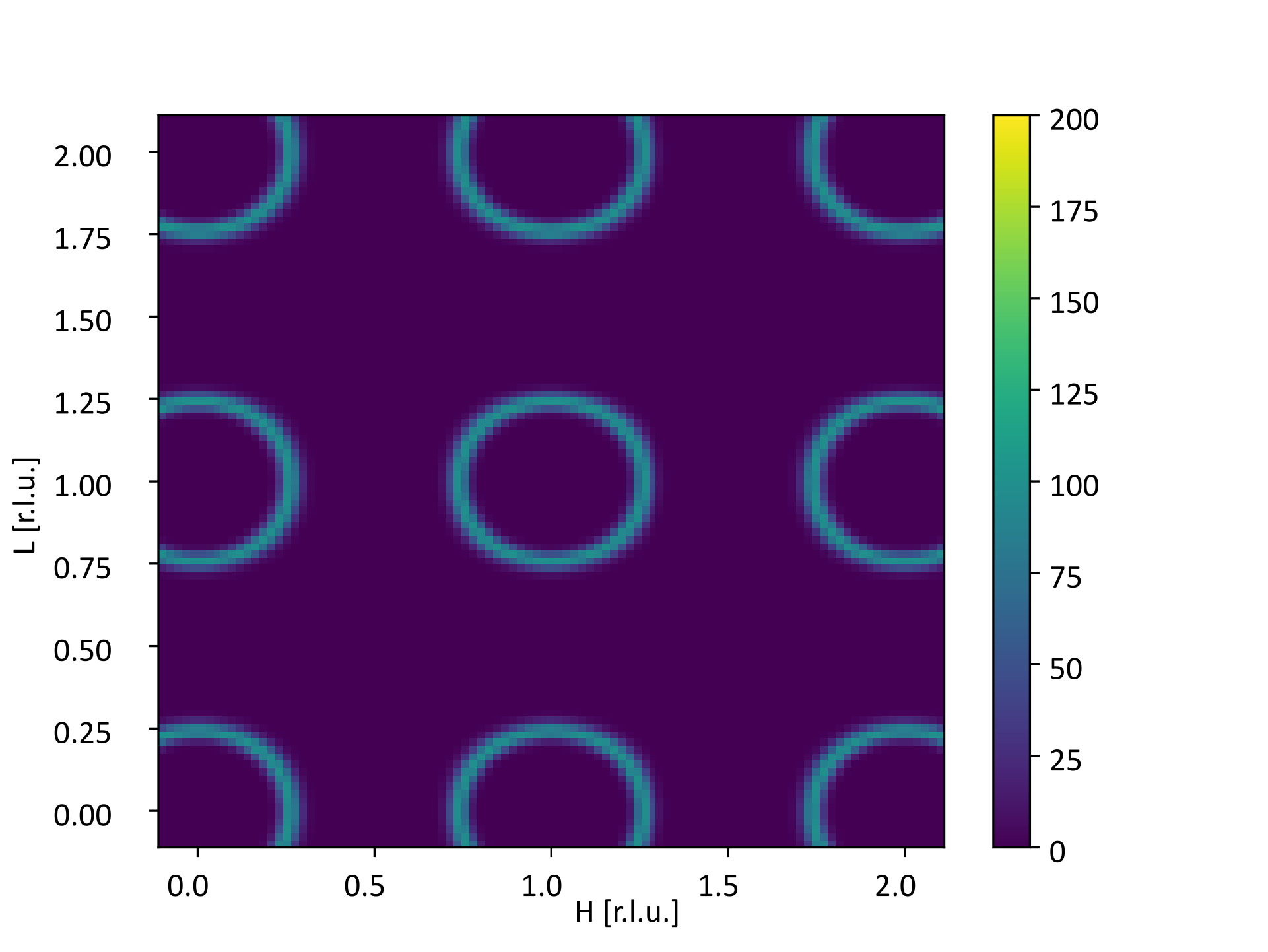}
      \caption{Energy cut ($E=4.55$ meV)}
      \label{fig:gt_data:Ecut}
    \end{subfigure}
    \caption{Ground truth neutron scattering intensity $\omega^2$($H$, $K$, $L$) as defined in Eq.~\eqref{eq:theoryMnF2}.}
    \label{fig:GroundTruthSyntheticMnF2}
\end{figure}

\begin{figure}
    \centering
    \begin{subfigure}{.5\textwidth}
      \centering
      \includegraphics[width=\linewidth]{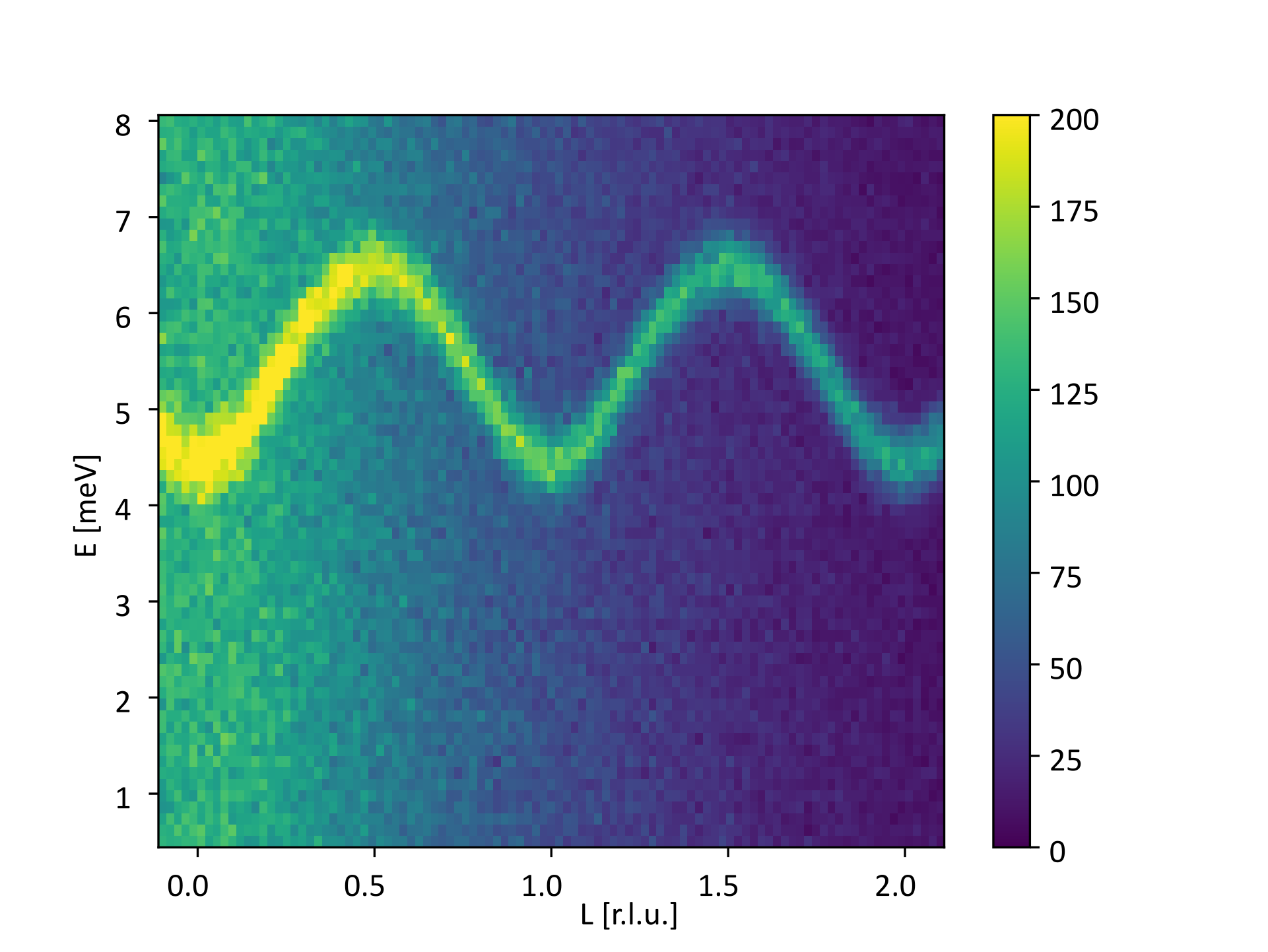}
      \caption{QvE plot ($H=0.25$)}
      \label{fig:noisy_data:QvE}
    \end{subfigure}%
    \begin{subfigure}{.5\textwidth}
      \centering
      \includegraphics[width=\linewidth]{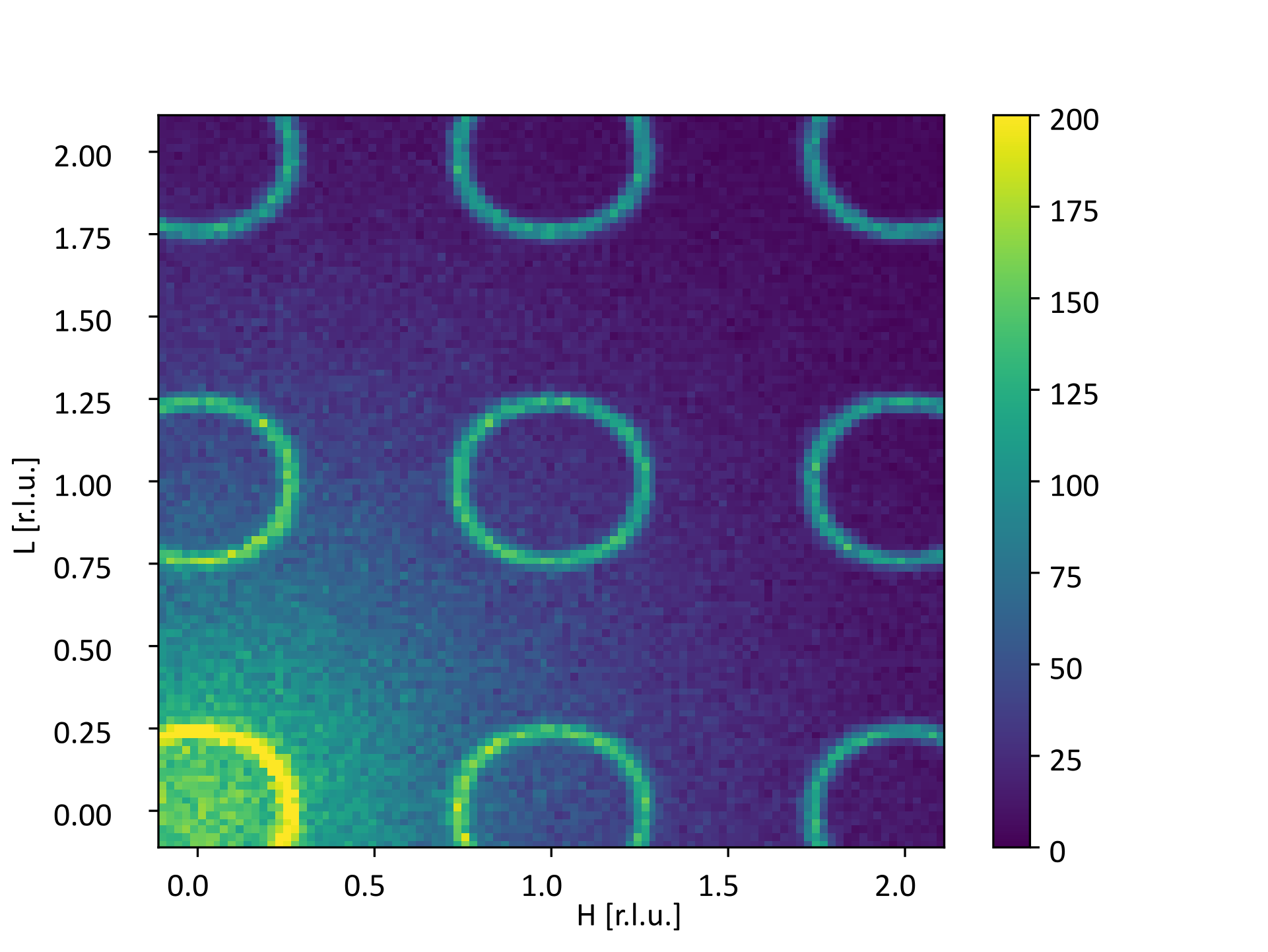}
      \caption{Energy cut ($E=4.55$ meV)}
      \label{fig:noisy_data:Ecut}
    \end{subfigure}
    \caption{Simulated neutron scattering intensity drawn from Poisson distribution with intensity function $\omega^2$ ($H$, $K$, $L$) and $\gamma =100.0$ and background function described in the main text, plotted (a) along ($Q,\omega$) and for a constant energy transfer of 2 meV.}
    \label{fig:SyntheticMnF2}
\end{figure}

\begin{figure}
    \centering
    \begin{subfigure}[t]{.49\linewidth}
      \centering
      \includegraphics[width=\linewidth]{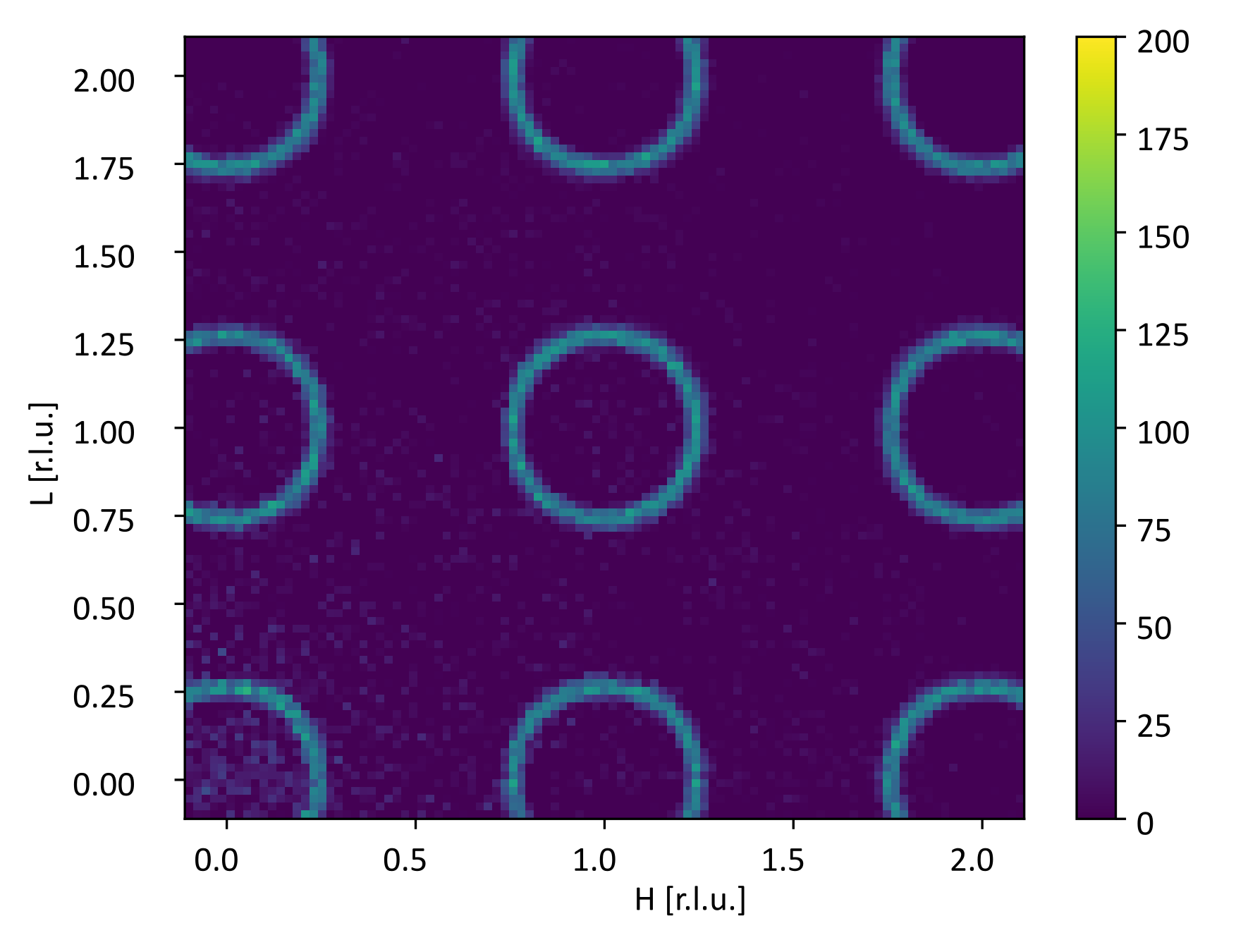}
      \caption{AMBER}
      \label{fig:synthetic:Ecut_amber}
    \end{subfigure}%
    \begin{subfigure}[t]{.49\linewidth}
      \centering
      \includegraphics[width=\linewidth]{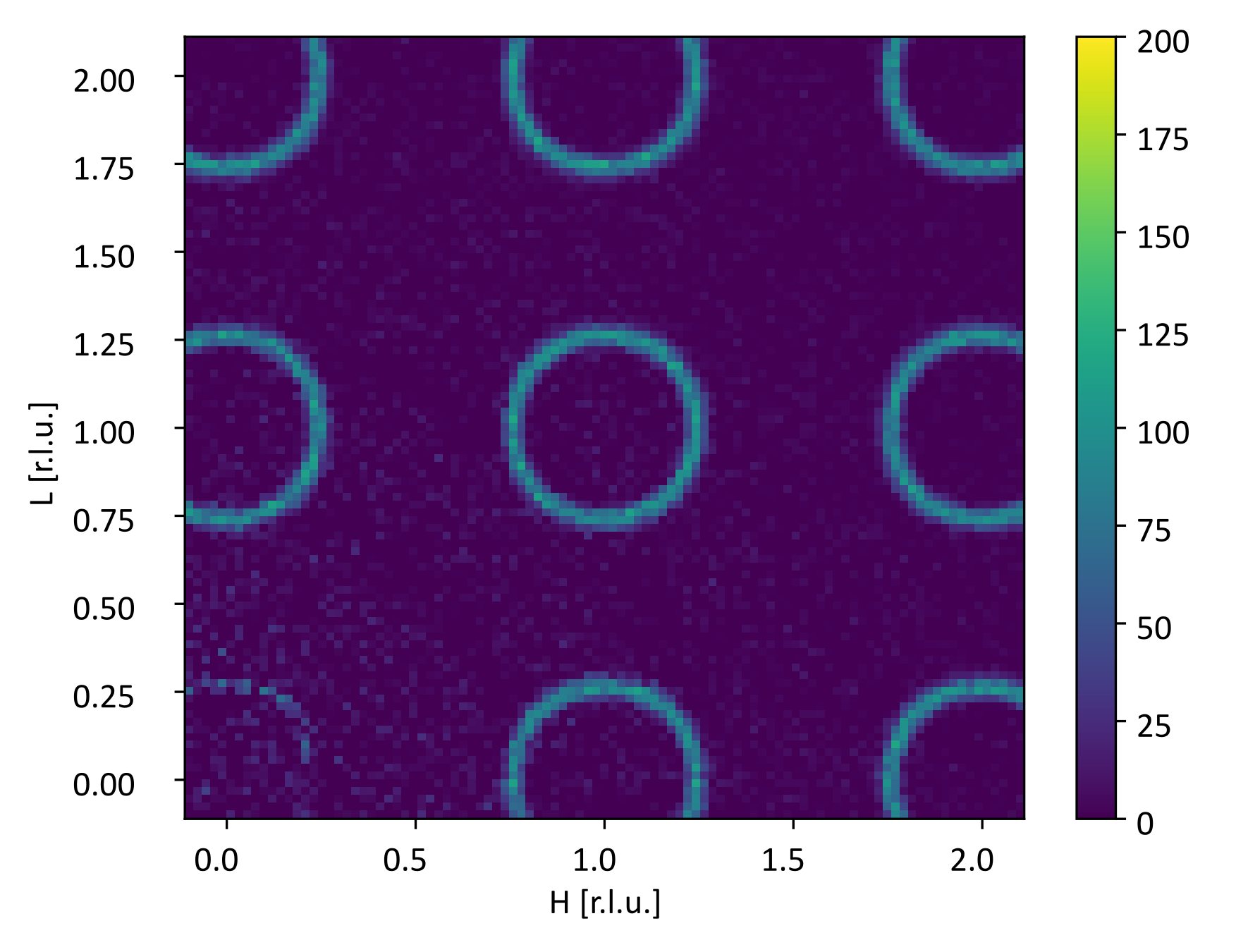}
      \caption{Baseline}
      \label{fig:synthetic:Ecut_med}
    \end{subfigure}
    \caption{Constant energy cut of synthetic data at $\Delta E = 4.55$ meV comparing our algorithm and the baseline segmentation method.}
    \label{fig:synthetic:Ecut}
\end{figure}

\begin{figure}
    \centering
    \begin{subfigure}[t]{.49\textwidth}
      \centering
      \includegraphics[width=\linewidth]{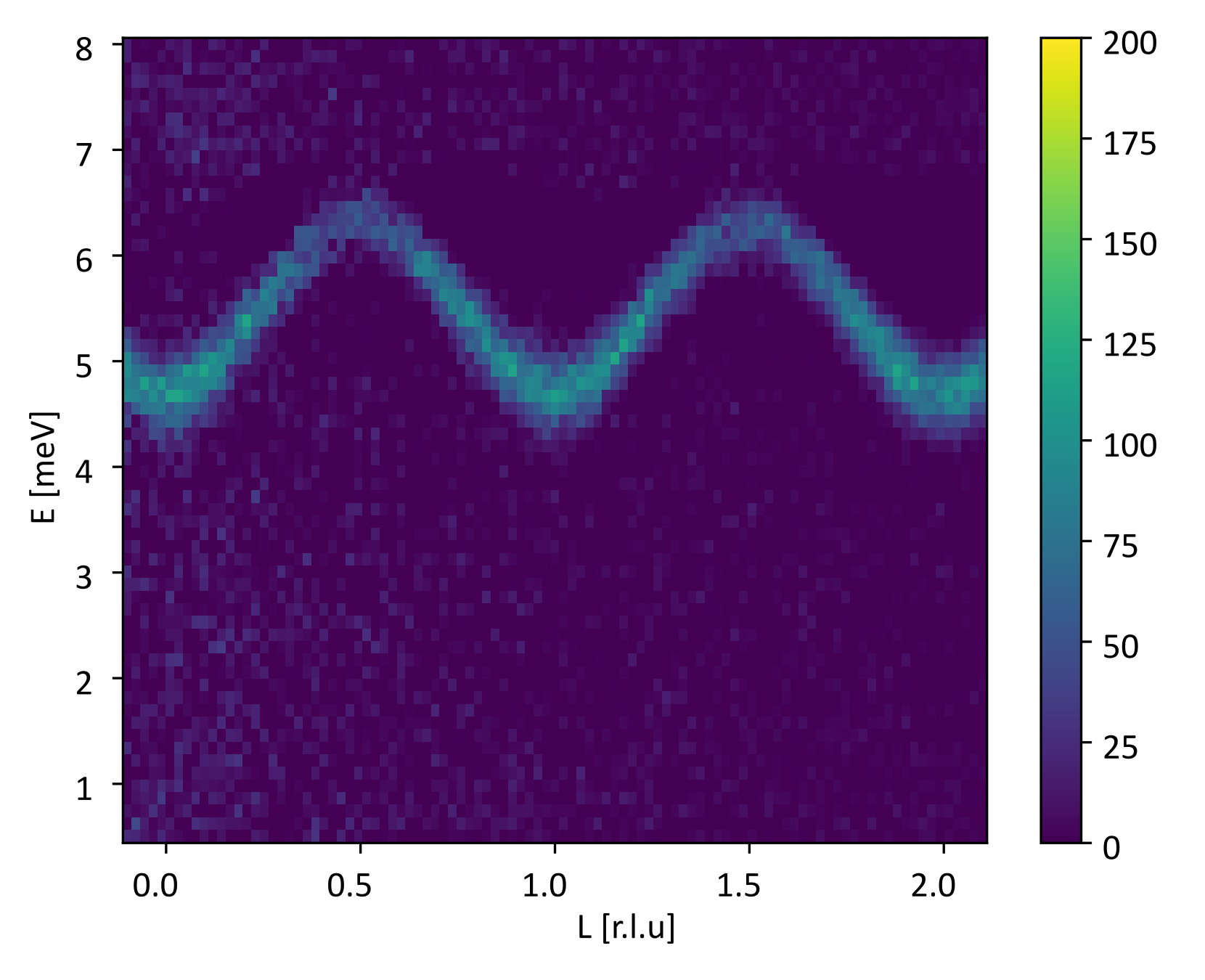}
      \caption{AMBER}
      \label{fig:synthetic:amber_QvE}
    \end{subfigure}%
    \begin{subfigure}[t]{.49\textwidth}
      \centering
      \includegraphics[width=\linewidth]{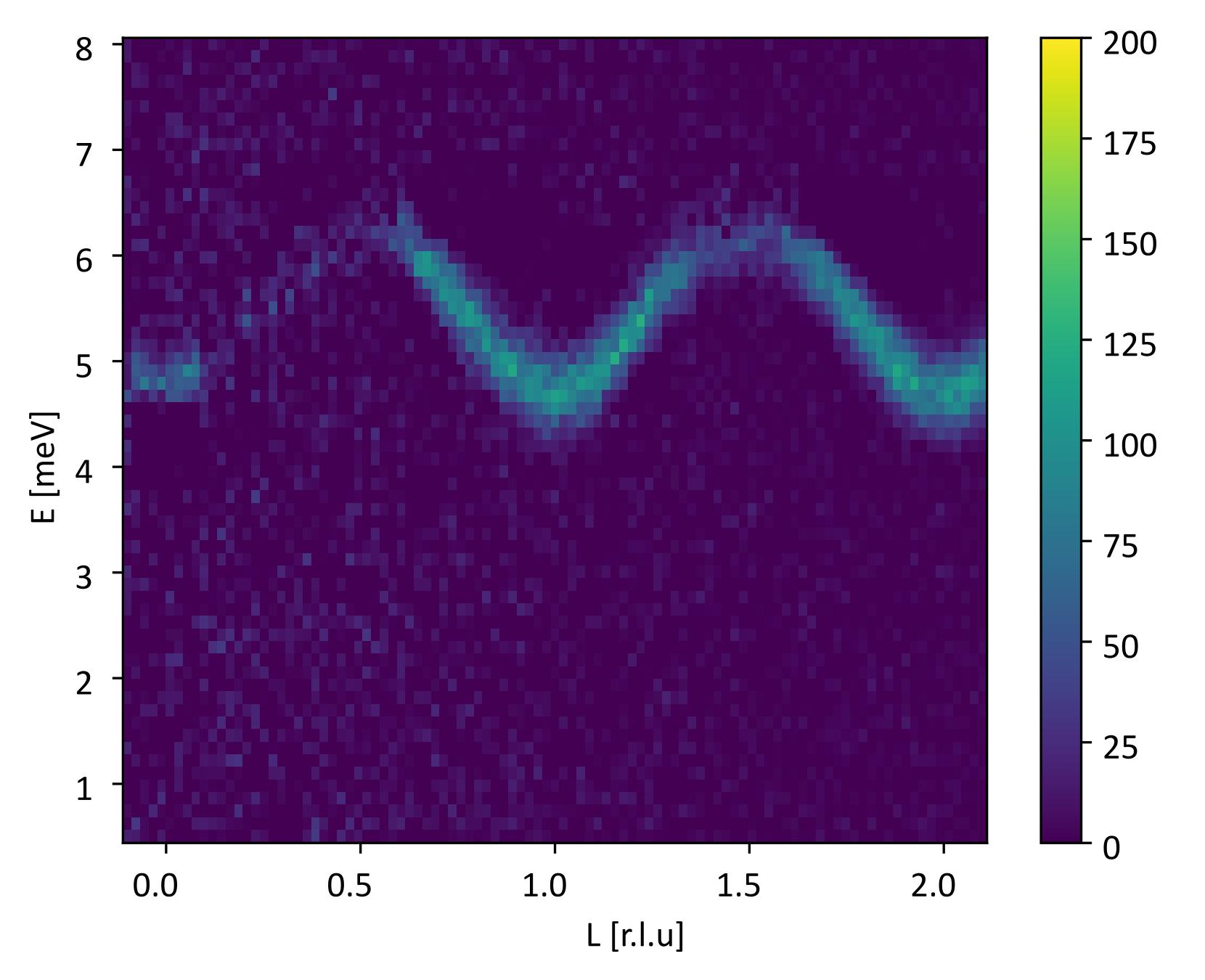}
      \caption{Baseline}
      \label{fig:synthetic:med_QvE}
    \end{subfigure}
    \caption{Plot of scattering intensity as a function of momentum and energy transfer between (0.25,0,-0.1) and (0.25,0,2.1) and from 0.5 to 8.0 meV comparing the ground truth, observed data, our algorithm and the baseline segmentation method.}
    \label{fig:synthetic:QvE}
\end{figure}

In this section we show two types of results. 
Firstly, we visualize the results of AMBER against the median-based baseline approach using intensity as function of energy and momentum transfer as well as momentum transfer in two directions at finite energy.  Secondly, we run the denoising algorithm for varying signal intensities, which leads to a varying signal-to-noise ratio (snr).
Notably, we ran the tuning procedure of Section~\ref{sub:hyperparam} with $\beta \in \{ 1.0,10.0,100.0,1000.0\}$. We evaluated the performances of \AMBER\ and the baseline approach by computing the Root Mean Squared Error (RMSE) with the ground truth signal. Figures~\ref{fig:synthetic:Ecut} and~\ref{fig:synthetic:QvE} show an energy cut visualization and a QvE visualization of the signal, respectively. One can observe that the baseline does not recover the signal in the region near to the origin ($H=0.0$,$L=0.0$).
We conclude that \AMBER\ captures the signal across all energy levels while the baseline approach fails to capture high energy level signal features.

The robustness of \AMBER\ has been tested by varying the amplitude of the signal using a scaling factor $\gamma$ in the intensity function, i.e., we define $\tilde{\omega}^2 = \gamma \omega^2$. 

We evaluated the performance of \AMBER\ against the baseline for 10 values of $\gamma$ equally spaced in $[1.0\mathrm{e}2; 1.0\mathrm{e}3]$ and we fix the amplitude of the background $A=100.0$.

\begin{figure}
    \centering
    \includegraphics[width=0.85\textwidth]{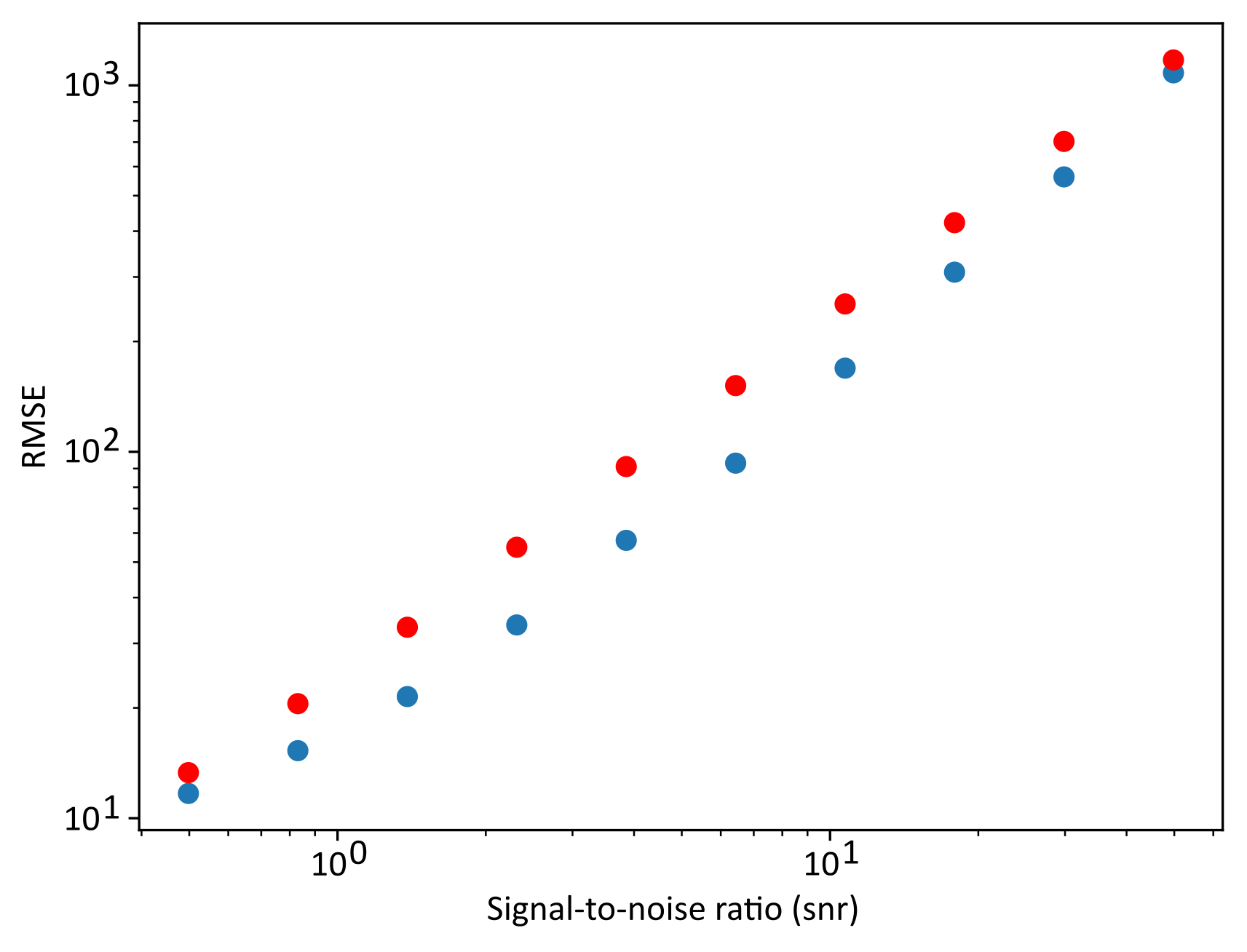}
    \caption{RMSE of \AMBER\ (blue) and the baseline approach (red) with respect to Signal-to-Noise ratio for each value of $\gamma$ in $[1.0\mathrm{e}2; 1.0\mathrm{e}3]$.}
    \label{fig:synthetic:snr}
\end{figure}

Figure~\ref{fig:synthetic:snr} displays the RMSE of both approaches with respect to S/N.
We conclude that our approach constantly outperforms the baseline approach. 


\FloatBarrier
\subsection{Experimental data - \texorpdfstring{MnF$_2$}{MnF2}}\label{App:MnF2} 

\begin{figure}
    \centering
    \begin{subfigure}[t]{.49\linewidth}
      \centering
      \includegraphics[width=\linewidth]{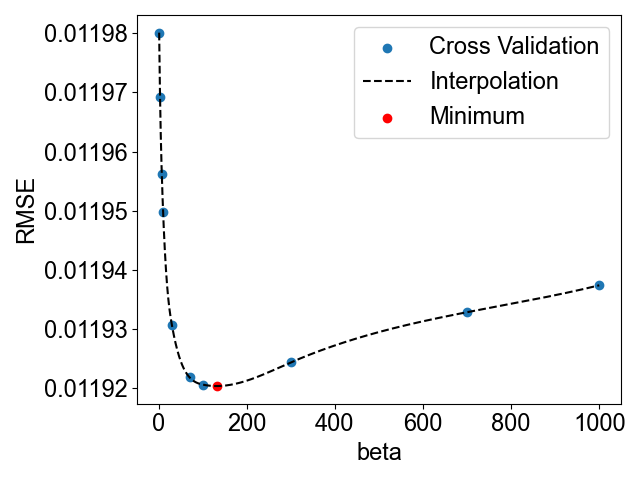}
      \caption{}\label{fig:MnF2BetaCrossValidation}
      \end{subfigure}\\
      \begin{subfigure}[t]{.49\linewidth}
      \centering
      \includegraphics[width=\linewidth]{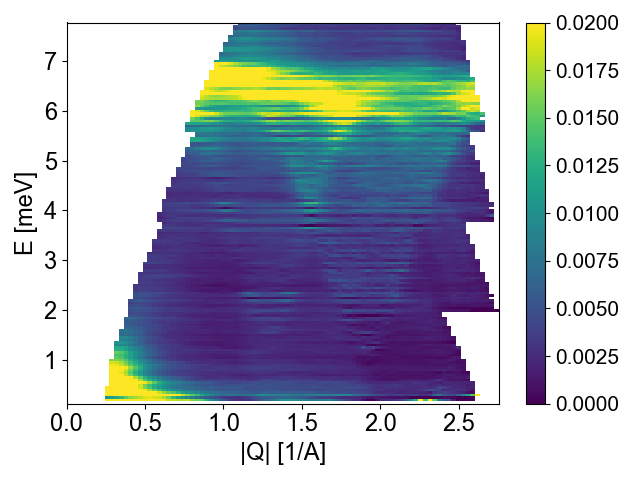}
      \caption{}\label{fig:AMBERPowderBackground}
      \end{subfigure}
      \begin{subfigure}[t]{.49\linewidth}
      \centering
      \includegraphics[width=\linewidth]{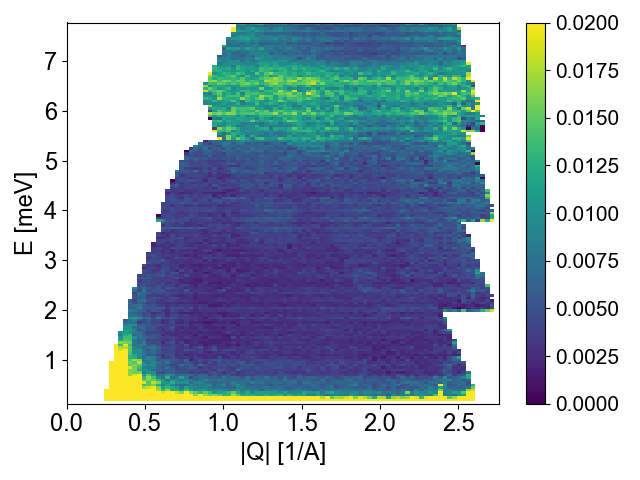}
      \caption{}\label{fig:MaskingPowderBackground}
      \end{subfigure}
    \caption{(a) RMSE as function of $\beta$ for the experimental MnF$_2$ data over-plotted with a quadratic spline. Minimum is found at $\beta$ = 131.5. (b) Background intensity predicted by \AMBER\ as function of $|Q|$ and energy. (c)  Background intensity predicted by the masking method as function of $|Q|$ and energy. The two backgrounds are very similar, except at an energy transfer of around 6.}
\end{figure}

In this section, we evaluate the performance of AMBER on data that have been acquired using the experimental setup of the CAMEA presented in sec.~\ref{sec:intro}.
The data were acquired at CAMEA in 2022 and consist of 32 data files. The measurement settings are tabulated in Tab.~\ref{tab:MnF2CAMEA}. The original data were acquired by measuring around 5 seconds per point but 5 times at the same position. In the data analysis, these individual counting intervals have been combined. The segmentation algorithm ran with the hyperparameters selected using the heuristic approach of Section~\ref{sub:hyperparam}, i.e. $\lambda$ was set as the Median Absolute Deviation scaled by $k = 1.468$, $\mu$ was set by the mean variance of the data along the energy direction and $\beta$ is found through a cross-validation shown in Fig.~\ref{fig:MnF2BetaCrossValidation}. We mention that this tuning procedure does not necessarily lead to the optimal estimated signal according to RMSE with signal ground truth but is quite close. This underlines that the tuning is heuristic. However, one can observe that we can get the best performance by using the optimal parameters obtained on the synthetic MnF$_2$ data. Since we have access to the simulation model, we can compute a background mask that is very close to the ground-truth. 

\begin{table}[!ht]
    \centering
    \begin{tabular}{c|c|c|c|c}
       Energy [meV]  &  A3 Range [deg] & A3 Steps & A4 [deg] & Time/Step [s]  \\ \hline
       5.37  & -45 - 75 & 62 & -40, -44, -74, -78 & 48.0 \\
       5.5  & -45 - 75 & 62 & -40, -44, -74, -78 & 48.0 \\
       7.13 & -48 - 74 & 62 & -40, -44, -70, -74 & 48.3 \\
       7.3 & -48 - 74 & 62 & -40, -44, -70, -74 & 50.0 \\
       8.97 & -51.2 - 70.8 & 62 & -40, -44, -60, -64 & 47.3 \\
       9.1 & -51.2 - 70.8 & 62 & -40, -44, -60, -64 & 48.2 \\
       10.77 & -54.4 - 67.6 & 62 & -40, -44, -50, -54 & 47.8 \\
       10.9 & -54.4 - 67.6 & 62 & -40, -44, -50, -54 & 48.1 \\
    \end{tabular}
    \caption{Measurement settings used for the acquisition of MnF$_2$ data at CAMEA. A scan is performed for each combination of energy and A$_4$, i.e. for e.g. an incoming energy of 5.5 meV a total of four scans have been performed. Thus, a total of 32 scans make up this data set.}
    \label{tab:MnF2CAMEA}
\end{table}
The bin sizes used for the background methods are 0.03 \AA\ along both $|Q|$-directions and 0.05 meV along the energy axis. The determined parameters of \AMBER\ were $\lambda = 0.004625$, $\mu =  0.06869$, and $\beta = 131.5$ found graphically with q = 0.9, see Fig.~\ref{fig:MnF2BetaCrossValidation}. The resulting background estimates are plotted in Figs. \ref{fig:AMBERPowderBackground} and \ref{fig:MaskingPowderBackground} for both the \AMBER\ method and the standard masking procedure, respectively. The masking method relies on a manual masking foreground signal and performing a powder averaging, see e.g. \cite{Facheris2022}. The two methods generate slightly different background estimates as function of energy and $|Q|$, but capture both the main background features, i.e. at low and high energy transfer.

\begin{figure}
    \centering
    \begin{subfigure}[t]{.49\linewidth}
      \centering
      \includegraphics[width=\linewidth]{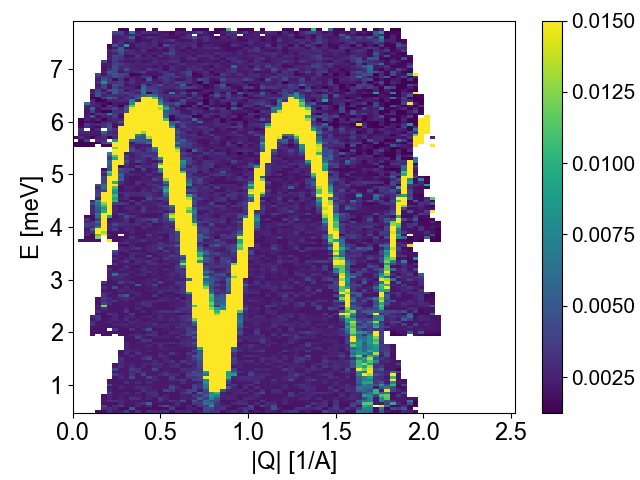}
      \caption{Masking Subtraction}
      \label{fig:instrument:QvE_gt}
    \end{subfigure}%
    \begin{subfigure}[t]{.49\linewidth}
      \centering
      \includegraphics[width=\linewidth]{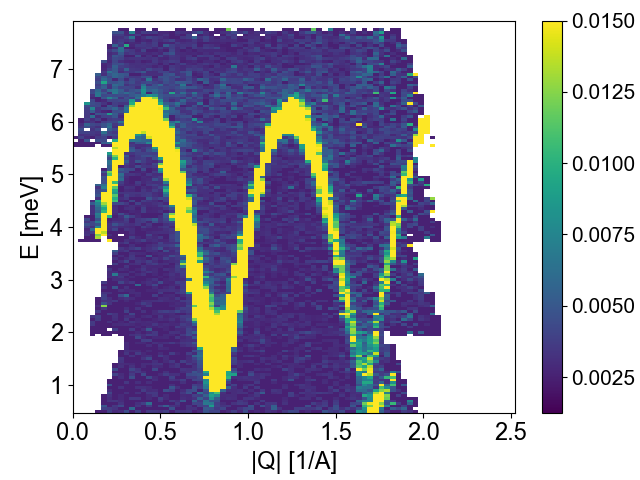}
      \caption{Observation}
      \label{fig:instrument:QvE_obs}
    \end{subfigure}\\%
    \begin{subfigure}[t]{.49\linewidth}
      \centering
      \includegraphics[width=\linewidth]{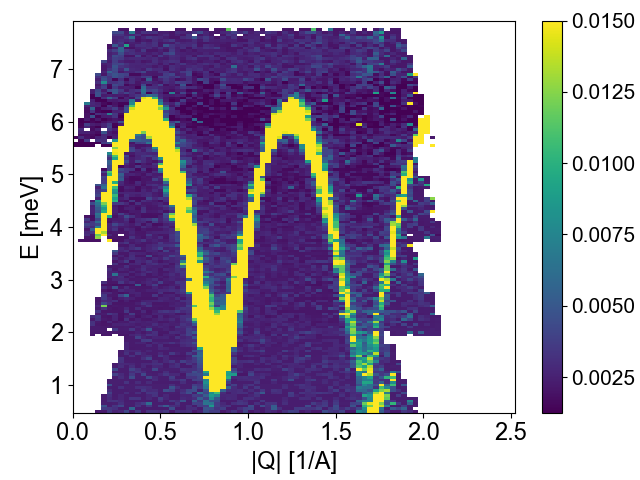}
      \caption{AMBER}
      \label{fig:instrument:QvE_algo}
    \end{subfigure}%
    \caption{Experimental scattering intensity for MnF$_2$ plotted as function of momentum and energy transfer between (-1,0,-1) and (-1,0,1) and 0 to 7 meV. Binning orthogonal to the cut is 0.05 \AA~and the same long the cut. In the energy direction, the bin size is 0.05 meV. (a) Scattering intensity background-subtracted using the masking method. (b) Observation. (c) Data subtracted using the \AMBER\ method.}
    \label{fig:instrument:QvE}
\end{figure}

\begin{figure}
    \centering
    \begin{subfigure}[t]{.49\linewidth}
      \centering
      \includegraphics[width=\linewidth]{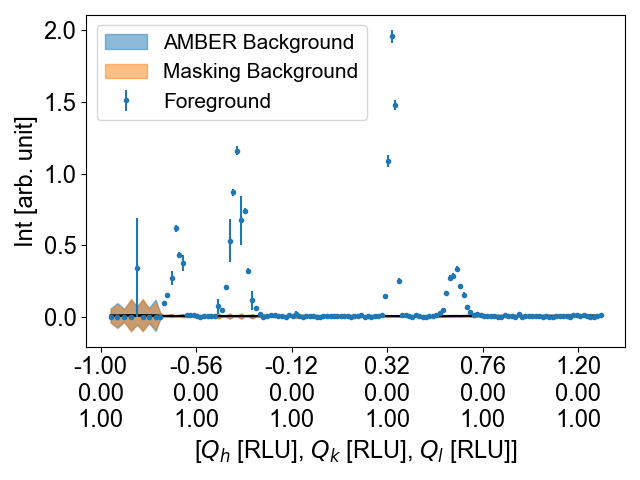}
      \caption{}
      \label{fig:instrument:Ecut_Masking}
    \end{subfigure}%
    \begin{subfigure}[t]{.49\linewidth}
      \centering
      \includegraphics[width=\linewidth]{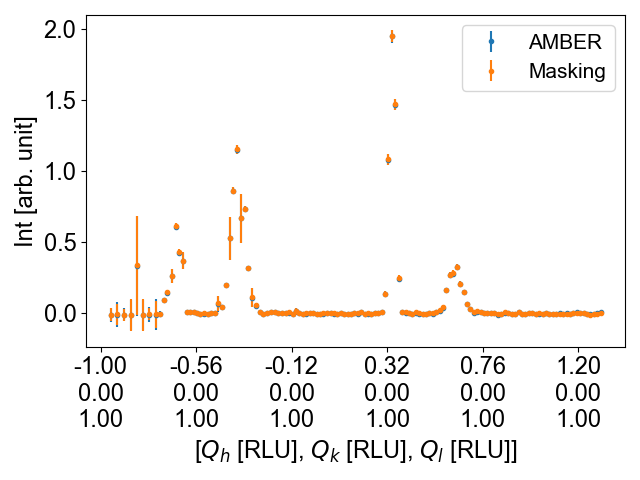}
      \caption{}
      \label{fig:instrument:Ecut}
    \end{subfigure}\\%
    \begin{subfigure}[t]{.49\linewidth}
      \centering
      \includegraphics[width=\linewidth]{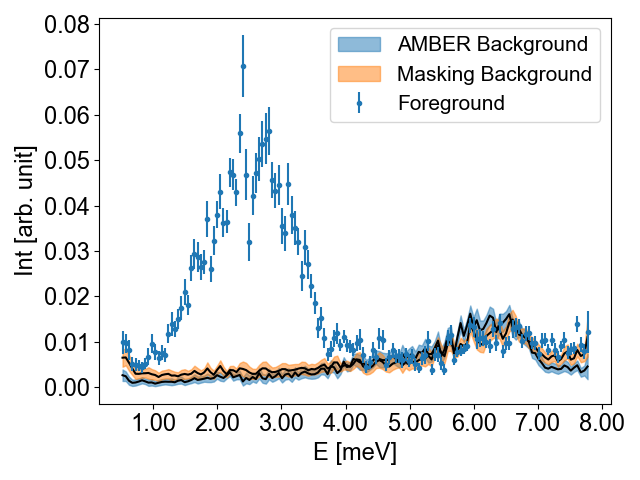}
      \caption{}
      \label{fig:instrument:Ecut_gt}
    \end{subfigure}%
    \begin{subfigure}[t]{.49\linewidth}
      \centering
      \includegraphics[width=\linewidth]{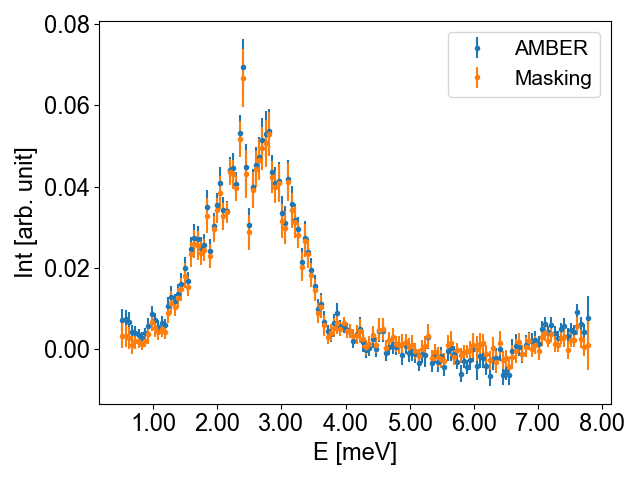}
      \caption{}
      \label{fig:instrument:Ecut_obs}
    \end{subfigure}\\%
    \caption{1D cut through MnF$_2$ data along energy in steps of 0.05 meV at $(0.9,0,1)$ integrated by 0.1 \AA. Only a slight difference in the background is visible at higher energy transfers. }
    \label{fig:MnF21D}
\end{figure}

The data are visualized in Fig.~\ref{fig:instrument:QvE} showing scattering intensity as a function of scattering vector along $(-1,0,L)$. 
The figure is split into three panels showcasing (a) the ground truth, i.e. using the standard background masking method (b) the experimental data, (c) background subtracted data utilizing \AMBER. 
In the case of MnF$_2$, the masking method removes signal  around predicted spin-wave positions. The three figures are very similar, with slight differences highlighted through 1D cuts along constant energy and constant $|Q|$. These are presented in Fig.~\ref{fig:MnF21D}. Figures (a) and (c) show the scattering intensity overplotted with the estimated background from the two methods, and (b) and (d) show the resulting background-subtracted intensities. For the constant energy cut, \AMBER\ and the masking method perform very similarly with barely any difference while for constant $|Q|$ \AMBER\ performs a slight over-subtraction at higher energies. 

This example shows that \AMBER\ is less eficient than the masking method for cases in which a perfect intensity masking can be performed. 

\FloatBarrier

\section{Derivation of coordinate descent}\label{App:Proof}
\label{app:r_star}

In this section, we explain how we derive the expressions in Eq.~\eqref{eq:coordinate_descent_Xupdate} and~\eqref{eq:coordinate_descent_Bupdate}.

Coordinate descent updates~Eq.\eqref{eq:coordinate_descent_Xupdate} and~\eqref{eq:coordinate_descent_Bupdate} are obtained by minimizing $\mathcal{L}$ with respect to to $X$ and $b$, respectively.
It turns out that these updates are equivalent to proximal gradient steps with

\begin{align*}
    \arg\!\min_{X} \mathcal{L} \left( X,b\right) &= \arg\!\min_{X} \frac{1}{2}\lVert Y-X-\mathcal{R}b\rVert_{2}^2 + \lambda \Vert X \Vert_1 + \frac{\mu}{2} \boldsymbol{1}_{n_x}^T X L_{\omega} X^T \boldsymbol{1}_{n_y} \\
    &= \arg\!\min_{X} \frac{1}{2}\lVert Y-\mathcal{R}b - X\rVert_{2}^2 + \lambda h(X) +\frac{\mu}{2} g(X) \\
    & = \mathrm{prox}_{\lambda h + \frac{\mu}{2} g} \left( Y-\mathcal{R}b \right) \\
\end{align*}
where $h(x) = \Vert x \Vert_1$ and $g(x) = \boldsymbol{1}_{n_x}^Tx L_{\omega} x^T \boldsymbol{1}_{n_y}$.
Therefore, we need to compute the proximal operator of the sum of a $l_1$ penalty and a quadratic norm. Thanks to known results on elastic net regularization~\cite{Trevor2005,Boyd2014}, we have $\mathrm{prox}_{\lambda h + \frac{\mu}{2} g} = \mathrm{prox}_{\frac{\mu}{2} g} \circ \mathrm{prox}_{\lambda h}$. 
\begin{align*}
    \arg\!\min_{X} \mathcal{L} \left( X,b\right) &= \mathrm{prox}_{\frac{\mu}{2} g} \circ \mathrm{prox}_{\lambda h} \left( Y-\mathcal{R}b \right) \\
    &= (I + \mu L_{\omega} )^{-1} \mathrm{prox}_{\lambda \Vert \cdot \Vert_1} \left( Y-\mathcal{R}b \right) \\
    &= (I + \mu L_{\omega} )^{-1} S_{\lambda} \left( Y-\mathcal{R}b \right) \\
\end{align*}
which achieves the proof of the coordinate descent update with respect to $X$.

Now we show how we derive Eq.~\eqref{eq:coordinate_descent_Bupdate}. To do so, we compute the gradient of $\mathcal{L}$ with respect to $b$. First we reformulate the loss function $\mathcal{L}$ as follows
\begin{align*}
    \mathcal{L}(X,b) 
    &= \frac{1}{2} \langle \mathcal{R} b,\mathcal{R} b \rangle + \frac{1}{2} \Vert Y-X \Vert_2^2 - \langle Y-X, \mathcal{R} b\rangle + \frac{\beta}{2} \langle b, b \rangle_{L_b} + \frac{\mu}{2} \boldsymbol{1}_{n_x}^T X L_{\omega} X^T \boldsymbol{1}_{n_y} \\
    &= \frac{1}{2} \langle b, \mathcal{R}^*\mathcal{R} b \rangle + \frac{1}{2} \Vert Y-X \Vert_2^2 - \langle b, \mathcal{R}^* (Y-X)\rangle + \frac{\beta}{2} \langle b, b \rangle_{L_b} + \frac{\mu}{2} \boldsymbol{1}_{n_x}^T X L_{\omega} X^T \boldsymbol{1}_{n_y},
\end{align*}
where we used the adjoint operator property. At the end of the section, we show how to compute $\mathcal{R}^*$.
Here, we derive the gradient of $\mathcal{L}$ with respect to $b$.

\begin{align*}
    \nabla_{b} \mathcal{L} (X,b) &= \mathcal{R}^*\mathcal{R} b - \mathcal{R}^*(Y-X) + \beta L_q b \\
    &= \left( \mathcal{R}^*\mathcal{R} + \beta L_q\right) b - \mathcal{R}^*(Y-X) 
\end{align*}
Therefore, by setting the gradient to zero, we obtain that $rg\!\min_{b} \mathcal{L} (X,b)= \left(\mathcal{R}^* \mathcal{R} + \beta L_q \right)^{-1}\mathcal{R}^*(Y-X)$, which is the update~\eqref{eq:coordinate_descent_Bupdate}.

\noindent We explain how to compute $\mathcal{R}$. We use the definition of Hermitian adjoint, i.e., $\langle \mathcal{R} u, v \rangle = \langle u, \mathcal{R}^* v \rangle$ for any $u \in \mathbb{R}^{n_q \times n_{\omega}}$ and $u \in \mathbb{R}^{n_x \times n_y \times n_{\omega}}$.
The left-hand side can be written
\begin{align*}
    \langle \mathcal{R} u, v \rangle &= \sum_{i,j,e} (\mathcal{R} u)_{i,j,e} v_{i,j,e} \\
    &= \sum_{e} \sum_{r} \sum_{(i,j) \in r} (\mathcal{R} u)_{i,j,e} v_{i,j,e} \\
    & = \sum_{r,e}  u_{r,e} \sum_{(i,j) \in r} v_{i,j,e} \\
    &= \langle u,  (\sum_{(i,j) \in r} v_{i,j,e})_{r,e} \rangle
\end{align*}
where the second line comes from the rewriting of the sum over the pixels in $\vec{Q}$ space as a sum over pixel in each radial bin $r$ in polar coordinates of the $\vec{Q}$ space. Hence, we can deduce that
\begin{align*}
    \mathcal{R}^* v = \left( \sum_{(i,j) \in r} v_{i,j,e}\right)_{r,e}
\end{align*}
Therefore, $\mathcal{R}^*$ is the operator that computes for each radial bin the sum across all angles.
It follows that $\mathrm{argmin}_{b} \mathcal{L}(X,b) = \left(\Gamma + \beta L_q \right)^{-1} \mathcal{R}^* (Y-X)$ where $\Gamma$ is a matrix whose elements are the number of measurements within each radial bin.
It achieves the proof.

\section{Energy-dependent parameters}\label{App:ParameterExtension}

The solution of Eq.~\ref{pb:minimization}sketched in \ref{sec:solution} discusses the hyperparameters in terms of scalars. However, an extension enables $\lambda$ and $\beta$ to vary along the energy direction, i.e. $\lambda \rightarrow \boldsymbol{\lambda}_e$ and $\beta\rightarrow\boldsymbol{\beta}_e$. This allows for more flexibility in dealing with varying levels of signal and noise but has the drawback of making the method more complicated and introduces additional free parameters. 
Naturally, $\mu$ can also be extended but his would lead to a non-compact formulation in the presented solution in ~\ref{sec:solution}. Hence, $\mu$ remains a scalar parameter.  

That is, we introduce the two subscripts $e$ and $q$,  
$\boldsymbol{\lambda} = (\lambda_e)_e$ and $\boldsymbol{\beta} = (\beta_{e})_e$.
where $e=1,\ldots, n_{\omega}$, $q=1, \ldots, n_q$, $i=1,\ldots, n_x$ and $j=1,\ldots, n_y$. 
We could also extend $\mu$ to $\boldsymbol{\mu} = (\mu_{i,j})_{i,j}$ but this would lead to a non-compact formulation in the presented solution in Section~\ref{sec:solution}. Hence, $\mu$ remains a scalar parameter.  

Thus, Eq.~\ref{pb:minimization} is reformulated to 
\begin{align*}
    \min_{X,b} &\frac{1}{2}\lVert Y-X-\mathcal{R}b\rVert_{2}^2+ \sum_{e=1}^{n_{\omega}} \boldsymbol{\lambda} \vert| X_e |\vert_{1} + \sum_{e=1}^{n_{\omega}}\frac{\boldsymbol{\beta}}{2} \mathrm{Tr} \left( b_e^T L_{b} b_e \right)+ \frac{\mu}{2} \boldsymbol{1}_{n_x}^TX^T L_{\omega} X \boldsymbol{1}_{n_y}.
\end{align*}
where "$\cdot$" is the component-wise product, i.e., $u \cdot v = \left( u_i v_i \right)_{i}$. We emphasize that this does not lead to changes in other parts of the algorithm. 

\newpage

\section*{Current executable software version}
\label{sec:Currentexecutablesoftwareversion}
\begin{table}[!ht]
\begin{tabular}{|l|p{4.3cm}|p{8.3cm}|}
\hline
\textbf{Nr.} & \textbf{(Executable) software metadata description} & \textbf{AMBER} \\
\hline
S1 & Current software version & {\color{red}1.0.1}\\
\hline
S2 & Permanent link to executables of this version  & N/A   \\
\hline
S3 & Legal Software License & Mozilla Public License 2.0 (MPL-2.0)\\
\hline
S4 & Computing platforms/Operating Systems & Linux, OS X, Microsoft Windows, Unix-like \\
\hline
S5 & Installation requirements \& dependencies & Python  \\
\hline
S6 & If available, link to user manual - if formally published include a reference to the publication in the reference list & \url{https://AMBER-ds4ms.readthedocs.io/} \\
\hline
S7 & Support email for questions & \url{jakob.lass@psi.ch}  \\
\hline
\end{tabular}
\caption{Software metadata (optional)}
\label{tab:SoftwaremetadataAppendig} 
\end{table}

\end{document}